\documentclass[useAMS,usenatbib]{mn2e}

\def\aap{AA}

\def\apjl{ApJL}

\def\mnras{MNRAS}
\def\apj{ApJ}

\def\aj{AJ}

\def\nat{Nat}
\def\ds{\displaystyle}

\def\Kv{{K}}
\def\RM{{R_{\rm M}}}
\def\RS{{R_\sigma}}

\usepackage{graphicx}
\usepackage{float}
\usepackage{amssymb}
\usepackage{amsfonts}
\usepackage{amsmath} 
\usepackage{color}

\title[Elliptical Galaxies: Simple Dynamical Models]{Dynamical Models
 of Elliptical Galaxies -- I. Simple Methods} 

\author[A. Agnello, N. W. Evans,
   A. J. Romanowsky]{A. Agnello$^{1}$\thanks{Email:
     aagnello@ast.cam.ac.uk, nwe@ast.cam.ac.uk, aaron.romanowsky@sjsu.edu},
   N. W. Evans$^{1}$, A. J. Romanowsky$^{2,3}$ \\ 
   $^{1}$ Institute of Astronomy, University of
   Cambridge, Madingley Road, Cambridge CB3 0HA, UK\\ 
   $^{2}$ Department of Physics and Astronomy, San Jos{\`e} State
   University, One Washington Square, San Jos{\`e}, CA 95192, USA\\ 
   $^{3}$ University of California Observatories, 1156 High Street, Santa
   Cruz, CA 95064, USA}

\begin{document}

\voffset-.6in

\date{Accepted . Received }

\pagerange{\pageref{firstpage}--\pageref{lastpage}} 

\maketitle

\label{firstpage}

\begin{abstract}
We study dynamical models for elliptical galaxies, deriving the
projected kinematic profiles in a form that is valid for general
surface brightness laws and (spherical) total mass profiles, without
the need for any explicit deprojection. We provide accurate
approximations of the line of sight and aperture-averaged velocity
dispersion profiles for galaxies with total mass density profiles with
slope near $-2$ and with modest velocity anisotropy using only single
or double integrals respectively. This is already sufficient to
recover many of the kinematic properties of nearby ellipticals.

As an application, we provide two different sets of mass estimators
for elliptical galaxies, based on either the velocity dispersion at a
location at or near the effective radius, or the aperture-averaged
velocity dispersion. In the large aperture (virial) limit, mass
estimators are naturally independent of anisotropy. The spherical mass
enclosed within the effective radius $R_{\rm e}$ can be estimated as
$2.4 R_{\rm e} \langle \sigma^{2}_{\rm p} \rangle/ G$, where $\langle
\sigma^2_{\rm p} \rangle$ is the average of the squared velocity
dispersion over a finite aperture. This formula does not depend on
assumptions such as mass-follows-light, and is a compromise between
the cases of small and large apertures sizes. Its general agreement
with results from other methods in the literature makes it a reliable
means to infer masses in the absence of detailed kinematic
information. If on the other hand the velocity dispersion profile is
available, tight mass estimates can be found that are independent of
the mass-model and anisotropy profile. In particular, for a de
Vaucouleurs surface brightness, the velocity dispersion measured at
$\approx 1 R_{\rm e}$ yields a tight mass estimate (with 10 \%
accuracy) at $\approx 3R_{\rm e}$ that is independent of the mass
model and the anisotropy profile. This allows us to probe the
importance of dark matter at radii where it dominates the mass budget
of galaxies.
 
Explicit formulae are given for small anisotropy, large radii and/or
power-law total densities. Motivated by recent observational claims,
we also discuss the issue of weak homology of elliptical galaxies,
emphasizing the interplay between morphology and orbital structure.
\end{abstract}

\section{Introduction}

Galaxies are known to contain both luminous and dark matter (DM). In
particular, DM haloes provide the seeds of galaxy formation, as
baryons cool and fall towards the centres of DM overdensities in
protoclusters, resulting eventually in the luminous, directly
observable components. Once gas is converted into stars, the assembly
of central objects proceeds via mergers \citep{cat11,joh12}.

Cosmological DM-only simulations offer predictions as to the shape,
density profile and typical mass of DM haloes \citep{nfw96}. However,
the buildup of baryonic matter affects the DM haloes in which they
assemble, through gravitational interaction between the luminous and
dark component.  When baryonic effects are included in the
simulations, these can transfer energy between the luminous and dark
components and alter the DM profile through different channels
 \citep{aba10,dic13}. In particular, in elliptical galaxies
 baryonic feedback \citep{dub13} and virialisation of the infalling
 material \citep{lac10} can produce a shallower density profile, whereas a slow
 mass build-up tends to steepen it \citep{blu86,lac10}.

When the assembly of central objects is studied with higher-resolution
and smaller-scale simulations, a set of prescriptions must be adopted
to quantify the importance of baryonic feedback, amount of
substructure and merging rates. These yield distinctive signatures on
the final state, in terms of size and mass of the stellar component as
well as DM content and density profile \citep{nip12,hil13,rem13}.

Then, investigating the DM profiles of observed galaxies provides
tests of galaxy formation scenarios. The task is simpler for late-type
galaxies, in which the orbits of the stars are generally
near-circular. In early-type galaxies, the role of the mass profile in
the observed kinematics is degenerate with the orbital distribution of
stars. This is commonly known as the \textit{mass-anisotropy
  degeneracy}, and constitutes the main obstacle to robust conclusions
on the dynamics of elliptical galaxies.  Equivalently, only the
projected observables (surface brightness and line-of-sight
velocities) are available, whereas the dynamics of these systems is
characterized by the deprojected, three-dimensional densities and
velocities.

The investigation of DM in elliptical galaxies usually relies on
techniques that construct three-dimensional models and compare their
projected properties to the observational data. This approach is
traditionally implemented via the Jeans equations governing the
velocity moments of the distribution function, adopting or relaxing
the approximation of spherical symmetry
\citep{ems94,ev94,cap06,cap13}. A more rigorous alternative
considers distribution functions and orbit modelling for the luminous
component \citep[and references
  therein]{sch79,ric84,ber94,ev94a,car95,kr05}, which has also the
advantage of encoding the whole kinematic information beyond the
second velocity moments \citep{mer93,ger98}.

When the kinematic information is averaged over some {spatial}
aperture, such as in integral-field or long-slit spectroscopy of
unresolved stellar populations, the importance of orbital structure is
reduced. Then, a theoretical framework that naturally encodes
aperture-averaging would put the stress on the adopted physical model,
rather than on the numerical details that are inherent in, for
example, orbit-based descriptions. Within the Jeans formalism in
spherical, the projected velocity dispersion $\sigma_{\rm p}$ follows
from the density and anisotropy profiles. \citet{mam05} provided
expressions of $\sigma^{2}_{\rm p}$ in terms of single integrals of
mass profile and luminosity density, for a set of simple anisotropy
models.  \citet{mam05a} reduced the expressions for aperture-averaged
velocity dispersions from triple integrals (usually shown in the
literature) to single ones in the isotropic case.  Here, we develop an
approach that operates just within the direct observables, in
particular the surface brightness profile rather than the luminosity
density. This has already been studied by \citet{aae13} in the context
of gravitational lensing by early-type galaxies. In this paper, we
extend our earlier formalism to include the role of anisotropy
explicitly within different models.

In Section 2, we present new formulae for line of sight and
aperture-averaged velocity dispersions. Within the approach followed
here, there is no need to perform any explicit or approximate
deprojection.  Section 3 provides simple explicit results, for
scale-free densities or modest anisotropy and/or large radii. We
compare our findings to empirical aperture corrections that are
commonly used elsewhere. We show that some structural properties (such
as kinematic profiles and typical masses, Figs~\ref{fig:sapCapp} and
~\ref{fig:mestReff}) of early-type galaxies can be understood by means
of simple models, perhaps even deceptively simple!  In Section 4, we
present different mass estimators based on our formalism and we
characterise the possible sources of error. We sum up our conclusions
in Section 5. The methods illustrated below are particularly useful in
the presence of noisy data (e.g. Paper II in this series) or poor
spatial resolution of the measured kinematics.

\section{Line-of-sight Kinematics}
\subsection{Preliminaries}
We consider spherical models, such that the velocity dispersion tensor
is diagonal in spherical coordinates $(r,\theta,\phi)$ and the only distinction is
between radial and tangential motions. Let the anisotropy profile be
written as
\begin{equation}
\beta(r)= 1-\frac{\langle
  v_{\theta}^{2}+v_{\varphi}^{2}\rangle}{2\langle v_{r}^{2}\rangle}\ .
\end{equation}
Then, the Jeans equation for supporting the stellar component with
luminosity density $\nu$ in a gravitational potential $\Phi$ is
\begin{equation}
{\partial (\nu\langle v_{r}^{2}\rangle) \over \partial r}
+\frac{2\beta\nu\langle v_{r}^{2}\rangle}{r}=-\nu{ \partial \Phi
  \over \partial r}.
\label{eq:jeans}
\end{equation}
Our models are stationary ($\partial_{t}\nu=\partial_{t}\Phi=0$), with
neither radial flows ($\langle v_{r}\rangle=0$) nor Hubble flow.
While this hypothesis is acceptable for the internal dynamics of
elliptical galaxies, the application of the Jeans equations to galaxy
clusters requires additional correction terms \citep{fal13}.

Using the shorthand
\begin{equation}
J_{\beta}(r,s)= \exp\left[{\int_{r}^{s}2\beta(r')\mathrm{d}r'/r'} \right]
\end{equation}
for the integrating factor, eq.~(\ref{eq:jeans}) is easily solved for
the radial velocity dispersion~\citep[e.g.,][]{vdm94,ae11}
\begin{equation}
\langle v_{r}^{2}\rangle = \frac{G}{\nu(r)}\int_{r}^{\infty}\frac{M(s)\nu(s)}{s^{2}}J_{\beta}(r,s)\mathrm{d}s\ ,
\label{eq:inte}
\end{equation}
where we have cast the radial force in terms of the enclosed mass
$M(r).$ Observations provide the projected velocity second moment
$\sigma_{\rm p}(R)$ at radius $R,$ which is given by
\begin{equation}
\Sigma \sigma_{\rm p}^{2}(R) =2\int_{R}^{\infty}\left(1-\beta(r)\frac{R^{2}}{r^{2}}\right)
\frac{\nu(r)\langle v_{r}^{2}\rangle r\mathrm{d}r}{\sqrt{r^{2}-R^{2}}}
\label{eq:proj}
\end{equation}
\citep{bin82}, where $\Sigma(R)$ is the surface brightness. The
luminosity density can be obtained from the surface brightness profile
via Abel deprojection,
\begin{equation}
\nu(r)=-\frac{1}{\pi}\int_{r}^{\infty}\frac{\partial_{R}(\Sigma(R))\mathrm{d}R}{\sqrt{R^{2}-r^{2}}}\ ,
\label{eq:depr}
\end{equation}
and inserted in eq. (\ref{eq:proj}). However, it can be useful to have
results that depend directly on the surface brightness profile,
without the need for explicit deprojection, integration of the Jeans
equations and re-projection. This contrasts with other methods,
which rely on numerical or approximate deprojections of fitting
profiles, and therefore is the subject of the following sections.

\begin{figure}[ht]
	\centering
        \includegraphics[width=0.45\textwidth]{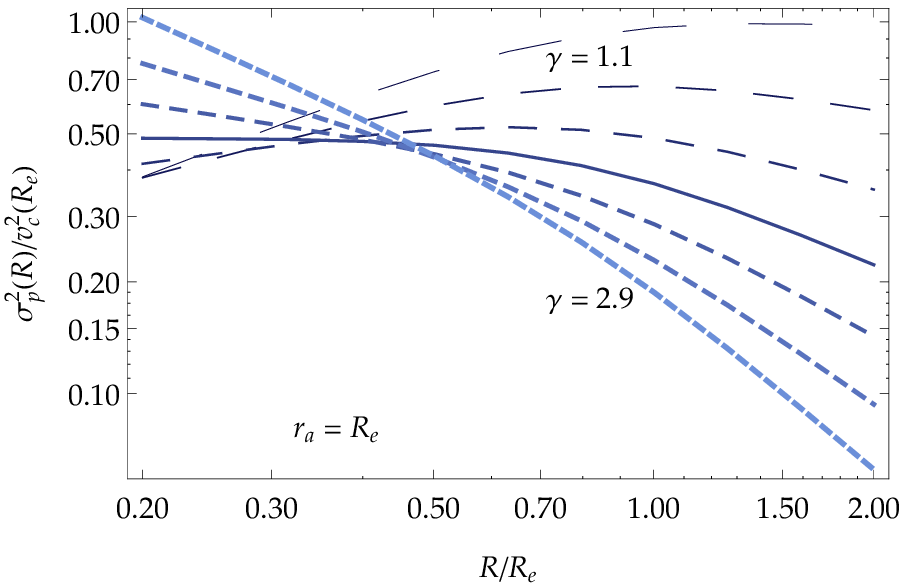}\\
        \includegraphics[width=0.45\textwidth]{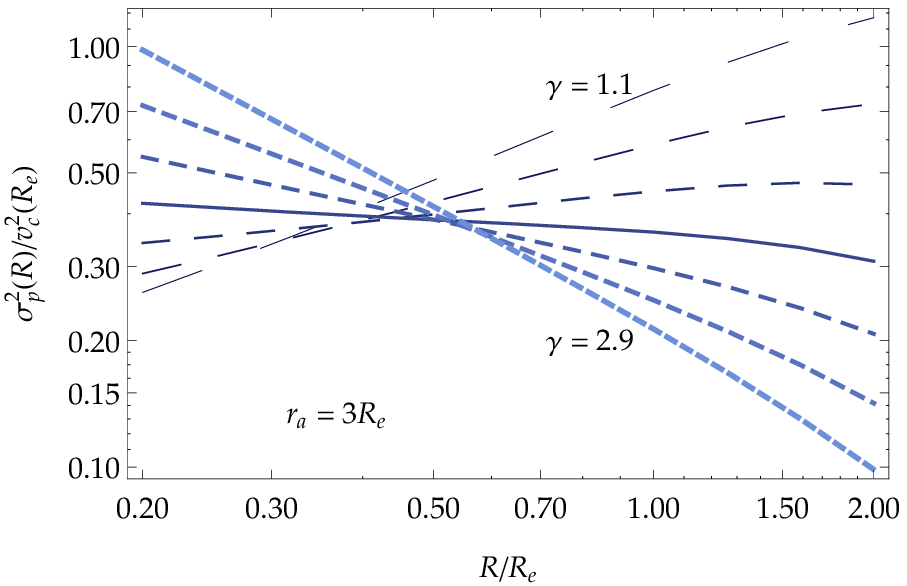}\\
        \includegraphics[width=0.45\textwidth]{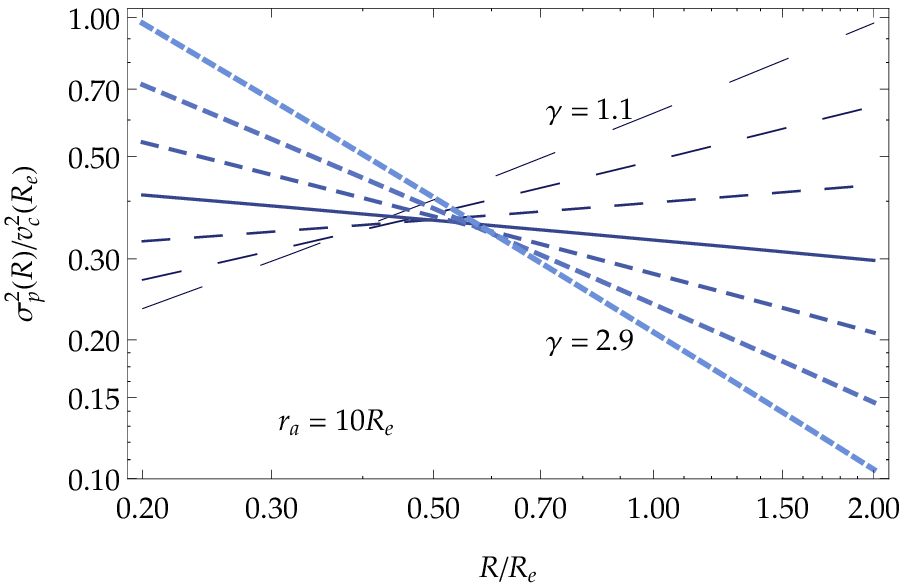}\\
\caption[Velocity dispersion profiles]{ \small{Profiles of squared
    projected velocity dispersion $\sigma^{2}_{\rm p}(R)$ rescaled to
    squared circular velocity $v^{2}_{\rm c}(R_{\rm e}),$ as a
    function of $R/R_{\rm e}$. Here, $\Sigma(R)$ is a de Vaucouleurs
    profile, the total density is $\rho_{\rm tot}\propto r^{-\gamma}$
    and the anisotropy profile is of Osipkov-Merritt form
    {(eq. \ref{eq:beta} with $\beta_{\infty}=1$)}.  The density
    exponent varies in steps of 0.3 between 1.1 (long-dashed, darkest,
    thinnest lines, labeled) to 2.9 (shortest-dash, clearest, thickest
    lines, labeled); full lines mark the flat rotation curve case of
    $\gamma=2.$ Different panels correspond to different values of
    {anisotropy radius $r_{\rm a}$} as in the legends.  Top: $r_{\rm
      a}=R_{\rm e};$ middle: $r_{\rm a}=3R_{\rm e};$ bottom: $r_{\rm
      a}=10R_{\rm e}.$ Pinch points, at which dependence on the
    adopted mass model is minimised, are present in each panel, but
    the location changes with anisotropy. }}
\label{fig:vdispR}
\end{figure}
\subsection{Line of Sight Velocity Dispersion Profiles}
Inserting eq.~(\ref{eq:inte}) in eq.~(\ref{eq:proj}), and exchanging
the orders of integration, an integration by parts leads to
\begin{equation}
\Sigma\sigma^{2}_{\rm p}(R) =2G\int_{R}^{\infty}\frac{\nu(r) M(r)}{r^{2}}
\left(\sqrt{r^{2}-R^{2}}+k_{\beta}(R,r)\right)\mathrm{d}r \ ,
 \label{eq:sigma}
\end{equation}
 where
\begin{equation}
k_{\beta }(R,x) =\int_{R}^{x}
 \frac{(2r^{2}-3R^{2})\beta(r)J_{\beta}(r,x)}{r \sqrt{r^{2}-R^{2}}} \mathrm{d}r\ .
\label{eq:kpdefn}
\end{equation}
The kernel $k_{\beta}(R,x)$ has already been expressed in analytical
form by \citet{mam05} for some particular choices of the anisotropy
profile.  Eq.~(\ref{eq:sigma}) gives the line-of-sight velocity
dispersion as a function of projected radius $R$. The dependence on
$\beta$ is separated out in the second integral on the right-hand
side. We can re-arrange this result explicitly in terms of the
observable stellar surface brightness $\Sigma$. First, we note the
useful identity
\begin{equation}
\frac{\mathrm{d}}{\mathrm{d}y}\int_{R}^{y}\frac{f(x,R)x}{\sqrt{y^{2}-x^{2}}}\mathrm{d}x=
y\int_{R}^{y}\frac{\partial_{x}f(x,R)}{\sqrt{y^{2}-x^{2}}}\mathrm{d}x\ ,
\label{eq:xy}
\end{equation}
which holds true if and only if $f(R,R)=0$ and provided the integrals
are well defined. Here, and elsewhere in this section, we defer the
technical details of proofs to Appendix A for the interested
reader. Inserting eq.~(\ref{eq:depr}) in eq.~(\ref{eq:sigma}),
integrating by parts and exploiting eq~(\ref{eq:xy}), we get in the
end
\begin{eqnarray}
\nonumber\Sigma\sigma_{\rm p}^{2}(R)=\frac{2G}{\pi}\int_{R}^{\infty}s\Sigma(s)
\int_{R}^{s}\frac{\partial_{r}\left(M(r)\sqrt{r^{2}-R^{2}}/r^{3}\right)}{\sqrt{s^{2}-r^{2}}}\mathrm{d}r\mathrm{d}s\ \\
+ \frac{2G}{\pi} \int_{R}^{\infty}s\Sigma(s)\int_{R}^{s}\frac{\partial_{r}
\left(M(r)k_{\beta}(R,r)/r^{3}\right)}{\sqrt{s^{2}-r^{2}}}\mathrm{d}r\mathrm{d}s\ .
\label{eq:projsigma}
\end{eqnarray}
This gives the line of sight velocity dispersion in terms of the
observable $\Sigma$ as well as model parameters such as the mass
$M(r)$ and anisotropy profile $\beta(r)$. It replaces the three
equations (\ref{eq:inte})-(\ref{eq:depr}), generalises equations (A15)
and (A16) of \citet{mam05} and obviates the need for explicit
projections and deprojections \citep[in eq. A8]{mam05}.  Isotropic
models ($\beta=0$) are all encoded in the first line, whilst the
second gives corrections for anisotropic models ($\beta\neq0$).

To make further progress, it is useful to introduce a two-parameter
family of anisotropy profiles
\begin{equation}
\beta(r)=\beta_{\infty}\frac{r^{2}}{r^{2}+r_{\rm a}^{2}}\ .
\label{eq:beta}
\end{equation}
This class of models allows us to examine systems where the anisotropy
changes gradually from isotropy at the center to a limiting value of
$\beta_{\infty}$ at large radii, as well as cases where the anisotropy is
fixed at a uniform value ($r_{\rm a} \rightarrow 0$).  The integrating
factor is simply
\begin{equation}
J_{\beta}(r,s)=\left(\frac{s^{2}+r_{\rm a}^{2}}{r^{2}+r_{\rm a}^{2}}\right)^{\beta_{\infty}}
\end{equation}
\citep[see][for the expression of $J_{\beta}$ for other anisotropy models]{mam13}.
 Although we will return to the generalised form ~(\ref{eq:beta}) in
Section ~\ref{sec:spec}, for the moment let us set $\beta_{\infty}=1$ so
that the models are strongly radially anisotropic at large radii.
Note that this corresponds to the ansatz introduced by Osipkov (1979)
and Merritt (1985). 

To gain insight, let us start with scale-free total densities,
$\rho_{\rm tot}\propto r^{-\gamma}.$ This choice is appropriate for
elliptical galaxies, at least within a few effective radii
\citep{tk04,mam05,gav07,hum10}. Fig.~\ref{fig:vdispR} shows the
typical behaviour of $\sigma_{\rm p}^{2}$ as a function of $R,$ for a
de Vaucouleurs luminous profile in different scale-free total
densities, having the same enclosed mass at the effective radius
$R_{\rm e}.$ The line-of-sight velocity dispersion has been normalised
to the circular velocity $v_{\rm c}(R_{\rm e})$ at the effective
radius to highlight the contribution from the mass profile rather than
from overall normalisations. Models with $\gamma>2$ have a falling
rotation curve and a declining velocity dispersion at all radii. When
$\gamma<2$ the velocity dispersion increases at small radii and
decreases slowly at large radii. The transition between these two
behaviours happens around $\gamma \approx 2$ (i.e. a flat rotation
curve), although the velocity dispersion profile is not exactly
flat. The exact value of the transition exponent, where
$\sigma_{p}(R)$ is almost uniform, varies depending on the structural
properties (e.g. S{\'e}rsic index and anisotropy).

More important than the shape of single velocity dispersion profiles
is the existence, for each chosen anisotropy, of a \textit{pinch
  radius} $R_{\sigma}$ where any dependence on the mass model is
minimal \citep{mam10,wol10}.  This location changes with anisotropy
$\beta$ (c.f., Fig.\ref{fig:vdispR}) and with the S{\'e}rsic index. In
particular, steeper profiles (lower S{\'e}rsic indices) produce a
smaller variation in $R_{\sigma}$ with $\beta.$ This fact can be
justified in the light of asymptotic behaviours at small $\beta$ or
large radii, which are discussed in Section \ref{sec:spec}; we will
exploit that in Section \ref{sect:makinpr} to construct a family of
mass estimators.

The behaviour of $\sigma_{\rm p}(R)$ with the effective radius is
controlled essentially by the circular velocity. If $R_{\rm e}$ is
increased, the overall normalisation decreases for $\gamma>2$ (as
$v_{\rm c}(R_{\rm e})\propto R_{\rm e}^{1-\gamma/2}$) and increases
for $\gamma<2.$ This means that, for a rising (declining) circular
velocity curve, increasing the effective radius will increase
(decrease) the overall magnitude of the velocity dispersion at fixed
$R/R_{\rm e}.$ This phenomenon is clear within scale-free total
densities and uniform anisotropy because, in this case, the only
available lengthscale is $R_{\rm e}$ and so we can expect
$\sigma^{2}_{\rm p}(R)$ to be modulated by $GM(R_{\rm e})/R_{\rm
  e}=R_{\rm e}^{2-\gamma}$ \citep[see, for example,][who give the
  exact solutions for scale-free tracers in scale-free total
  densities]{dek05}.

More elaborate mass models, exhibiting different power-law regimes in
different regions, can be understood in terms of the kinematic
profiles shown here. For example, a Navarro-Frenk-White density
$\rho_{\rm tot}\propto r^{-1}(1+r/r_{\rm s})^{-2}$ produces a line of
sight dispersion profile that is approximated by the one with
$\gamma\approx 1$ at small radii and $\gamma\rightarrow3$ at large
radii, provided $\Sigma(R)$ declines fast enough with $R.$ However, in
most cases, eq.~(\ref{eq:projsigma}) allows for an analytic evaluation
of the inner integral giving the mass-kernel, without any need for the
approximation of scale-free total densities.

\begin{figure}
	\centering
        \includegraphics[width=0.45\textwidth]{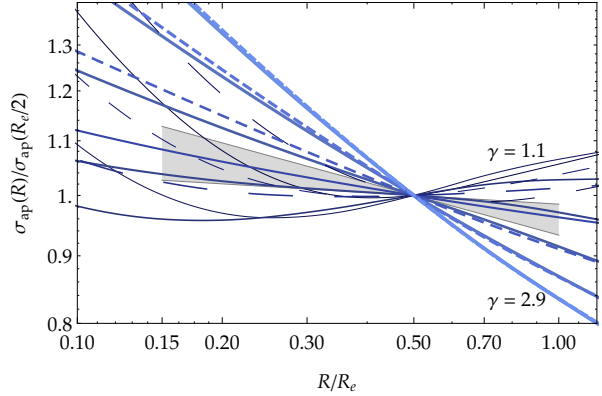}
\caption[Aperture-averaged velocity dispersions]{ \small{Line-of-sight
    velocity dispersion, averaged over an aperture of radius $R,$ as a
    function of $R/R_{\rm e}$ for a de Vaucouleurs luminosity profile
    in scale-free total mass densities, with exponent $\gamma$ ranging
    in steps of 0.3 from 1.1 (darkest, thinnest lines, labeled) to 2.9
    (lightest, thickest lines, labeled). The curves are
      computed using eq.~(\ref{eq:apsigma}). Every profile has been
    rescaled to the aperture-averaged velocity dispersion within
    $R_{\rm e}/2.$ Full lines: Osipkov-Merritt anisotropic models with
    $\beta(R_{\rm e})=1/2;$ dashed lines: isotropic models. The
    grey-shaded region shows the empirical relation $\sigma_{\rm
      ap}(R)\propto R^{-b},$ with $b=0.066\pm0.034$ (Cappellari et
    al. 2006).}}
\label{fig:sapCapp}
\end{figure}

\subsection{Aperture-averaged Velocity Dispersions}\label{sect:apavkin}

In practice, kinematics are measured over some aperture and blurred by
a point-spread function. Then, the quantity to be compared to
observations is the radial average
\begin{equation}
\sigma^{2}_{\rm ap}(R)
\equiv\frac{2\pi\int_{0}^{R}s\Sigma(s)\sigma_{\rm
    p}^{2}(s)\mathrm{d}s}{L(R)}\ ,
\end{equation}
with
\begin{equation}
L(R)=2\pi\int_{0}^{R}s\Sigma(s)\mathrm{d}s\ 
\label{eq:enclum}
\end{equation}
being the projected luminosity within $R.$ Averages within
radial annuli or slits can be derived from these formulae by means of
straightforward manipulations.

The triple integrals can be rearranged to express the
aperture-averaged velocity dispersion as a sum of three terms (see
Appendix A)
\begin{eqnarray}
\label{eq:apavps}
\sigma^{2}_{\rm ap}(R) & =& \frac{4\pi G}{3L(R)}
\left(\int_{0}^{\infty}M(r)\nu(r)r\mathrm{d}r\right. \nonumber\\
&-&\int_{R}^{\infty}M(r)\nu(r)\frac{(r^{2}-R^{2})^{3/2}}{r^{2}}\mathrm{d}r \\
&+&3\left. R^{2}\int_{R}^{\infty} \frac{M(r)\nu(r)}{r^{2}}
Z_{\beta}(R,r)\mathrm{d}r\right)\ \nonumber
\end{eqnarray}
where we have used the shorthand
\begin{equation}
Z_{\beta}(R,y)=\int_{R}^{y}J_{\beta}(r,y)\beta(r)\sqrt{r^{2}-R^{2}}{\mathrm{d}r\over r}\ 
\label{eq:defZ}
\end{equation}
The first line gives the virial limit, the second one provides
aperture corrections for $\beta=0$, while the third one expands to the
case of anisotropy $\beta\neq0$. Without the third line, this equation
is equivalent to the isotropic results of \citet{mam05}.  For
computational purposes, it is useful to replace the stellar density
$\nu$ in eq.~(\ref{eq:apavps}) with the stellar surface brightness
$\Sigma$ to obtain
\begin{eqnarray}
\label{eq:apsigma}
\sigma^{2}_{\rm ap}(R) & = & \frac{4G}{3  L(R)}
\left[\int_{0}^{\infty}\Sigma(s)s\int_{0}^{s}\frac{4\pi\rho_{\rm
      tot}(r)r^{2}}{\sqrt{s^{2}-r^{2}}}\mathrm{d}r\mathrm{d}s
  \right. \\
 \nonumber &-& \int_{R}^{\infty}\Sigma(s)s\int_{R}^{s}\frac{\partial_{r}\left(M(r)
 (r^{2}\!-\!R^{2})^{3/2}/r^{3}\right)}{\sqrt{s^{2}\!-\! r^{2}}}\mathrm{d}r\mathrm{d}s
 \\  \nonumber &+& \left. 3R^{2}\int_{R}^{\infty}\Sigma(s)s\int_{R}^{s}\frac{\partial_{r}\left(M(r)
 Z_{\beta}(R,r)/r^{3}\right)}{\sqrt{s^{2}\!-\! r^{2}}}\mathrm{d}r\mathrm{d}s
  \right]
\end{eqnarray}
The aperture-averaged velocity dispersion $\sigma^{2}_{\rm ap}(R)$ is
the outcome of two factors. The first is the mass model: as expected,
higher masses correspond to higher velocity dispersions at fixed
effective radius $R_{\mathrm{e}}.$ The second is the anisotropy, which
enters only in the last term of eq~(\ref{eq:apsigma}) and whose effect
on the velocity dispersion has the same sign as $\beta$. This means
that the uncertainties on the mass modelling due to observational
errors on the measured velocity dispersions can be decoupled from the
systematic uncertainties that are encoded in $\beta$
\citep[e.g.][]{koo09,aae13}.  The same remarks hold here for the
overall mass normalisation and behaviour with $R_{\rm e}.$

Fig.~\ref{fig:sapCapp} shows the behaviour of aperture-averaged
velocity dispersions $\sigma^{2}_{\rm ap}(R)$
scaled to the values at ${R_{\rm e}}/{2}$ in two cases --
namely, an Osipkov-Merritt profile with $\beta(R_{\rm
 e})=\tfrac{1}{2}$ and an isotropic model with $\beta=0$ everywhere.
 The choice of $R_{\rm e}/2$ is used solely to make comparisons
 with other work \citep{cap06} more immediate.
 
In general, models with $\gamma\lesssim2$ predict an averaged velocity
dispersion with a minimum at aperture radii between $R_{\rm e}/3$ and
$R_{\rm e}/2,$ increasing at both small and large apertures, whereas
steeper models produce a monotonically decreasing profile. The median
of the grey-shaded region in Fig.~\ref{fig:sapCapp}, corresponding to
the empirical relation $\sigma^{2}_{\rm ap}(R)\propto R^{-0.066}$
\citep{cap06}, is hardly distinguishable from a model with a de
Vaucouleurs luminous profile, a perfectly flat rotation curve and
$\beta=0.$ Models with $\beta(R_{\rm e})=\tfrac{1}{2}$ (full lines)
require slightly steeper density profiles to fit the grey band,
approximately $\gamma=2.1\pm0.1$. This small modulation of $\gamma$
with anisotropy suggests that, over lengthscales that are comparable
to the effective radius, {nearby} elliptical galaxies show weak
homology -- in the sense that their dynamical properties are
consistent with a total density scaling like $r^{-2}$ and just modest
radial anisotropy.

However, the median behaviour at radii $R_{\rm e}/2\lesssim R\lesssim
R_{\rm e}$ is not necessarily indicative of the density profile of
  single systems, especially over larger lengthscales. Analysis of
the hot X-ray gas in early-type galaxies by \cite{hum10} supports the
approximation of a scale-free total mass profile out to large radii,
but the relative exponent varies appreciably over their
sample. \citet{koo09} studied the density exponent $\gamma$ in 58
galaxies in the SLACS sample \citep{bol06}. The typical density
exponent from gravitational lensing, estimated by means of global
scaling relations over the whole sample, is in the interval
$\gamma_{l}=(2.03\pm0.07).$ On the other hand, on a galaxy-by-galaxy
basis the most likely density exponents occupy a much wider range,
with larger intrinsic uncertainties. The behaviour of $\gamma$ in
individual galaxies and the mean exponent $\gamma_{l}$ derived by
scaling relations over the whole sample are not directly related to
one another. Then, considerable care should be taken when the dynamics
of individual galaxies is studied, as to avoid the \textit{ecological
  fallacy} of exporting ensemble correlations at the individual
level. If the DM content at large radii is studied, simple analyses
enforcing $\gamma\approx 2$ may bias the inferred DM masses,
automatically favouring the values resulting from a flat rotation
curve.
 
The kinematic and photometric properties of individual galaxies can
deviate appreciably from the simple, average behaviour illustrated
above. In fact, the collection of profiles shown in \citet{cap06}, if
interpreted in terms of the models shown in Fig.~\ref{fig:sapCapp},
spans the whole range $1\lesssim\gamma\lesssim3$ and $r_{\rm a}\gtrsim
R_{\rm e}.$ In general, there is no guarantee that individual systems
are isotropic or that $\gamma=2$.  Moreover, the morphology of
individual galaxies can vary within the S{\'e}rsic family of
profiles~\citep{deV48,ser68}
\begin{equation}
\Sigma(R)=\Sigma_{0}\ \mathrm{exp}\left[-b_n (R/R_{\rm
    e})^{1/n}\right]\ ,
\label{eq:sers}
\end{equation}
where $b_n$ is defined such that $R_{\rm e}$ encloses half of the
total luminosity. A convenient expression of $b_{n}$ in $n$ has been
provided by \citet{cio99}. The light profiles of some elliptical
galaxies can be better fitted by S{\'e}rsic models with an index
substantially different from the de Vaucouleurs value $n =4.$ That
said, the assumption of weak homology can be taken as a first
approximation to infer properties of the mass profile within $R_{\rm
  e},$ before more detailed analyses are undertaken.


\section{Asymptotic Results}
\label{sec:spec}

\subsection{Line of Sight Velocity Dispersion Profiles}

A convenient aspect of the Jeans formalism is that eqs (\ref{eq:inte})
and (\ref{eq:proj}) involve information only from radii larger than
the upper limits of integration \citep[see e.g.,][]{vdm94,mam05}. In
particular, if the stellar density decays fast enough (which is always
the case for elliptical galaxies in practice), the dominant
contribution to the integrals is from radii just slightly greater than
the lower extremes of integration.  This turns out to be useful in
practice when handling the effects of anisotropy, since we just need
to consider the anisotropy profile and the mass $M(r)$ near the radii
of interest.

We will now analyse some applications of eq.~(\ref{eq:projsigma}). To this
end, we return to the generalisation of the Osipkov-Merritt anisotropy
profile given in eq~(\ref{eq:beta}). With this choice of $\beta$, the
kernel $k_{\beta}$ is:
\begin{eqnarray}
\nonumber k_{\beta}(R,r) &=& -\beta(R)\left(\frac{r_{\rm a}^{2}+r^{2}}
{r_{\rm a}^{2}+R^{2}}\right)^{\beta_{\infty}}\sqrt{r^{2}-R^{2}}\\
 &\times&
\left[F\left(\frac{1}{2},z\right)+\frac{2(1-r^{2}/R^{2})}{3}F\left(\frac{3}{2},z\right)\right]\ ,
\label{eq:hyperg}
\end{eqnarray}
where $z=(R^{2}-r^{2})/(r_{\rm a}^{2}+R^{2})$ and $F(a,z)$ is the
hypergeometric function $_{2}F_{1}\left(a,1+\beta_{\infty},a+1,z\right).$
Appendix B lists the special cases of $\beta_{\infty}=1,\tfrac{1}{2}$ and
$r_{\rm a}=0$.

For any surface brightness law, the kinematic profile is given by a
double integral where $\Sigma(R)$ is modulated by a kernel that
depends just on the potential chosen. The function $k_{\beta}$ can be
expanded in powers of $(r^{2}-R^{2})^{1/2}$ and the expansion
to first order is
\begin{equation}
k_{\beta }(R,r)\ \sim\ -\beta(R)\sqrt{r^{2}-R^{2}}+...\ .
\label{eq:kp}
\end{equation} 
If $\nu(r),$ and hence $\Sigma(R),$ decay fast enough with
radius $R,$ the next orders in the expansion can be neglected in a
first approximation.  If this is the case, the kinematic profile can
be obtained by neglecting the second line in eq.~(\ref{eq:projsigma})
and multiplying the first line by $1-\beta(R).$ This is useful for
obtaining asymptotic results at small and large radii.

An interesting class of results at small and large radii is provided
by scale-free densities, $\rho_{\rm
  tot}(r)=\rho_{0}(r/r_{0})^{-\gamma}.$ At small radii, we can rely
on the hypothesis of mild anisotropy. First, observations of nearby
elliptical galaxies \citep{ger01,cap06} show little or no departure from
isotropy inside $R_{\rm e}.$ Second, just a mild degree of anisotropy
is generally allowed in these systems by reasons of physical
consistency \citep{cio09}. This means that (see Appendix A for details)
\begin{equation}
\displaystyle
\frac{\Sigma(R)\sigma_{\rm p}^{2}(R)}{1\!-\!\beta(R)}= \frac{4\pi G\rho_{0}r_{0}^{\gamma}}{3\!-\!\gamma}
\int_{R}^{\infty} s^{1-\gamma}\Sigma(s)g_{\rm p}\left({R \over s},\gamma\right)\mathrm{d}s\ ,
\label{eq:plps}
\end{equation}
where
\begin{equation}
g_{\rm p}(x,\gamma)=\frac{1}{\pi}\int_{x^{2}}^{1}\frac{t^{-\gamma/2-1}
\left[(1-\gamma)t+\gamma x^{2}\right]}{\sqrt{t-x^{2}}\sqrt{1-t}}\mathrm{d}t\ .
\label{eq:gpex}
\end{equation}
The kernel $g_{\rm p}$ can be expressed as a combination of
hypergeometric functions and can be easily expanded in powers
of $x$.  An excellent approximation\footnote{This holds with
    $\lesssim0.3\%$ relative accuracy near the effective radius and
    $\approx 1\%$ at very small radii.} for $x\lesssim 1$ is
\begin{eqnarray}
\nonumber g_{\rm p}(x,\gamma) & \sim & \ 1 +{\gamma \over 2}(x-1)+{\gamma \over 8}\left(1-{\gamma \over 2}\right)(x-1)^{2}\\
 & + & \frac{\gamma^{2}(\gamma^{2}-4)}{96}(x-1)^{3}\ .
\label{eq:gpapp}
\end{eqnarray}
The result $g_{\rm p}=x$ for $\gamma=2$ (flat rotation curve) is
exact. 

At large radii, we cannot necessarily assume $|\beta|\ll 1.$ However, we can
approximate the kernel in the integrals for $y\gtrsim R$ as done above
in eq.~(\ref{eq:kp}). Higher orders only become important for high
values of $y,$ where the integrand is suppressed by the declining
$\Sigma(y).$ Also, we can use the asymptotic limit $\beta\rightarrow
\beta_{\infty}$ for the anisotropy profile. For $r\gg R,$ the kernel
$k_{\beta }$ grows at most linearly with $r$ (which happens when
$\beta_{\infty}=1$).  For $\beta\sim\beta_\infty$ and $r\gtrsim R$, we have
\begin{equation}
 \frac{k_{\beta}(R,r)}{\beta_{\infty}R}\ \sim\ -\left[1-\left(1-\tfrac{2}{3}\beta_{\infty}\right)\delta^{2}
 +\tfrac{3}{5}\left(1-\tfrac{2}{3}\beta_{\infty}\right)^{2}\delta^{4}
 \right]\delta +...
\end{equation}
 where $\delta=\sqrt{r^{2}/R^{2}-1}.$ This allows us to write
 $\sigma_{\rm p}^{2}$ at large radii as a single quadrature involving
 the tracer density $\nu,$ the mass profile $M$ and a sum of
 elementary functions \citep[cf][]{mam05}.  Alternatively, the result
 can be stated in terms of the surface brightness, exploiting
 eq.~(\ref{eq:xy}) in the same manner as done to derive
 eq.~(\ref{eq:projsigma}).
 
In particular, for scale-free total densities, the velocity dispersion
profile at large radii is asymptotically
\begin{eqnarray}
\nonumber \Sigma(R)\sigma_{\rm p}^{2}(R)\ &\sim & \frac{4\pi G\rho_{0}r_{0}^{\gamma}}{3\!-\!\gamma}
 \int_{R}^{\infty}s^{1-\gamma}\Sigma(s)\left[(1\!-\!\beta_{\infty})g_{\rm p}\left({R \over s},\gamma\right)\right.\\
 &+&\left.\beta_{\infty}\left(1\!-\!\tfrac{2}{3}\beta_{\infty}\right)h_{\rm p}\left({R \over s},\gamma\right)\right]\mathrm{d}y\ ,
\end{eqnarray}
 with
\begin{equation}
h_{\rm p}(x,\gamma)=\frac{x^{-2}}{\pi}\int_{x^{2}}^{1}\frac{t^{-\gamma/2-1}\sqrt{t-x^{2}}[\gamma x^{2}+(3-\gamma)t]}{\sqrt{1-t}}\ \mathrm{d}t\ ,
\end{equation} 
having retained just the two terms in equation (24).  The kernel
$h_{\rm p}$ can be expanded as
\begin{equation*}
\nonumber h_{\rm p} \sim\
\begin{cases} 
\frac{\displaystyle \Gamma\left((3-\gamma)/2\right)}{\displaystyle \sqrt{\pi}\Gamma\left(2-\gamma/2\right)}\left((3-\gamma)x^{-2}-3(1-\gamma/2)\right)\\
\nonumber \qquad\qquad\qquad +\mathcal{O}(x^{3-\gamma}), & x\ll 1\\
   \\
3(1\!-\!x)\!-\!\frac{3}{4}(-6\!+\!\gamma)(1-x)^{2}\\
\nonumber \qquad\qquad\qquad -\frac{\displaystyle 96-\gamma(14+\gamma)}{\displaystyle 16}(1-x)^{3}, & x\lesssim 1.
\end{cases}
\end{equation*}
In the important flat rotation curve case ($\gamma=2$), the result
\begin{equation}
h_{\rm p}(x,2)=x^{-2}\left(1-x^{3}\right)
\end{equation}
holds at all orders. 

\subsection{Aperture-averaged Velocity Dispersions}

For small anisotropy or large aperture radii, eq.~(\ref{eq:apsigma})
admits a simple approximation -- namely, we may again suppress the
third addendum and multiply the second one by $1-\beta(R)$. As a check
on our working, we note that for large values of aperture radius $R$,
we must recover the virial limit exploited elsewhere
\citep{ae12,ae12b}.

We again derive the results for mildly anisotropic systems in
scale-free total densities. Starting with eq.~(\ref{eq:apsigma}),
using the approximation for small $\beta$ and exchanging orders of
integration as before, we obtain:
\begin{eqnarray}
\nonumber \sigma^{2}_{\rm ap}(R)=\frac{16\pi G\rho_{0}r_{0}^{\gamma}}{3(3-\gamma)L(R)}
 \times \left(k_{\rm ap}(0,\gamma)\int_{0}^{\infty}\Sigma(s)s^{3-\gamma}\mathrm{d}s\right.\ \\
  -\left. (1-\beta(R))\int_{R}^{\infty}\Sigma(s)s^{3-\gamma}k_{\rm ap}(R/s,\gamma)\mathrm{d}s\right)
\label{eq:iso}
\end{eqnarray}
(cf Agnello et al. 2013). Again, the kernel
\begin{eqnarray}
\nonumber k_{\rm ap}(x,\gamma) =
(4-\gamma)\int_{x}^{1}\sqrt{\frac{u^{2}-x^{2}}{1-u^{2}}}u^{3-\gamma}\mathrm{d}u\\
 +(\gamma-1)x^{2}\int_{x}^{1}\sqrt{\frac{u^{2}-x^{2}}{1-u^{2}}}u^{1-\gamma}\mathrm{d}u
\end{eqnarray}
can be easily expanded in powers of $x:$
\begin{equation}
k_{\rm ap}(x,\gamma)\sim
\begin{cases}
\frac{\ds \sqrt{\pi}\Gamma((5-\gamma)/2)}{\ds \Gamma(2-\gamma/2)}\times
\left[1-\frac{\ds (1-\gamma/2)x^{2}}{\ds 1-\gamma/3}\right. & \\
\left.\qquad\qquad\qquad\quad -\frac{\ds \gamma(1-\gamma/2)x^{4}}{\ds 4(1-\gamma/3)}\right] & x\ll 1,\\
  &  \\
\frac{\ds 3\pi}{\ds 2}(1-x)-\frac{\ds 3\pi}{\ds 8}(2+\gamma)(1-x)^{2} & \\
\qquad\qquad\qquad\quad  +\frac{\ds \pi\gamma(10-\gamma)}{\ds 32}(1-x)^{3} & x\lesssim 1.
\end{cases}
\end{equation}
The result
\begin{equation}
k_{\rm ap}(x,2)=\frac{\pi}{2}(1-x^{3})
\label{eq:28}
\end{equation}
is exact. As a specific example, when we use the anisotropy
law~(\ref{eq:beta}), we find that our simple asymptotic approximation
is excellent for $r_{\rm a}\gtrsim 3R_{\rm e}$. In fact, provided the
models are reasonably close to the flat rotation curve case
($1.5\lesssim\gamma\lesssim 2.5$), it performs remarkably well even
when $r_{\rm a}=R_{\rm e}$.

The trick for reducing the eqs~(\ref{eq:projsigma}) and
~(\ref{eq:apsigma}) for the line of sight and aperture-averaged
velocity dispersions is of wider applicability. In each case, the
integrals over stellar surface density and total mass are greatly
simplified with little loss of accuracy when the anisotropy dependent
term is discarded and the previous term multiplied by $1
-\beta(R)$. The same trick can also be applied to eqs~(\ref{eq:sigma})
and (\ref{eq:apavps}) for which the integrals are written in terms of
the stellar density and total mass, if so desired.  This then gives
single integrals to express both line of sight and aperture-averaged
velocity dispersions for arbitrary velocity anisotropy profiles,
generalising results obtained by \cite{mam05a,mam05} in special cases.

Finally, we give in Appendix B formulae for the line of sight and
aperture-averaged velocity dispersion valid for small anisotropy
and/or large radii without the assumption of power-law densities. The
formulae are simpler than eqs~(\ref{eq:projsigma}) and
~(\ref{eq:apsigma}), as they involve just the total density $\rho_{\rm
  tot}$ and integrals over the surface brightness $\Sigma$.

\section{Mass estimators}

In the previous sections, we have seen how the line of sight
kinematics can be computed, starting from the mass profile $M(r)$ and
a choice of anisotropy profile $\beta.$ Now we ask a complementary
question: given the \textit{measured} kinematics, what is the best
inference that we can make on the mass profile?

The dimensional scaling $\sigma^{2}_{\rm p}\propto GM/R$ between the
second moment of line of sight velocities, enclosed mass and size is
evident in the Jeans formalism (e.g.  eqs \ref{eq:projsigma} and
\ref{eq:apsigma}). The inverse passage from $\sigma^{2}_{\rm p}(R)$ to
$M(r)$ is possible when $\beta(r)$ is given and the kinematic profile
is measured with sufficient accuracy \citep{mam10}. However, these
conditions are hardly satisfied in practice. Also, observational data
are often not sufficient to constrain all the parameters in the mass
profile. So, the problem of relating the measured kinematics to mass
estimates is often simplified to finding relations of the kind
\begin{equation}
\frac{GM(\RM)}{\RM}\ \equiv v_{\rm c}^{2}(\RM)\ =\ \Kv\sigma^{2}(\RS),
\label{eq:GMR}
\end{equation}
such that any model-dependence is minimal at the locations $\RS,\RM,$
while the parameter $\Kv$ is to be determined. Here, $\sigma^{2}(R)$
could be either the line of sight velocity second moment
(eq. \ref{eq:projsigma}) or the one averaged inside an aperture of
radius $R$ (eq. \ref{eq:apavps}), whilst $v_{\rm c}(R)$ denotes the
circular velocity at radius $R.$

This issue has been already tackled in a piecemeal manner in the
literature.  \cite{Ill76} derived a formula for constant mass-to-light
ratio models with a de Vaucouleurs profile. The total mass $M$ is
\begin{equation}
M(\infty) \approx {8.5R_{\rm e}\over G} \langle\sigma^{2}_{\rm p}\rangle\ ,
\label{eq:ill}
\end{equation}
where $\langle\sigma^{2}_{\rm p}\rangle$ is the average value of the
squared line of sight velocity dispersion.

\citet{cap06} studied 25 galaxies in the SAURON survey \citep{bac01},
by means of Jeans equations and orbit-based models. Their analyses
suggest a general trend
\begin{equation}
M(\infty) \approx {5R_{\rm e}\over G} \sigma^{2}_{\rm ap}(R_{\rm e})\ ,
\label{eq:cap5}
\end{equation}
where again $M(\infty)$ is the total mass and $\langle\sigma^{2}_{\rm
  p}\rangle (R_{\rm e})$ is the luminosity-weighted average over one
effective radius.  The formula holds if there is a negligible DM
fraction within the effective radius or, alternatively, if the light
traces mass. \citet{cap06} argued that accounting for an extended DM
halo would change the proportionality coefficient in
eq.~(\ref{eq:cap5}) by $\approx12\%$. This result is calibrated
against diverse, high spatial-resolution kinematic profiles (out to
$R_{\rm e}$), but its simplicity makes it useful for application to
galaxies for which any kinematic information is not as rich. However,
the main drawback of eqs~(\ref{eq:ill}) and (\ref{eq:cap5}) is the
assumption of a mass-follows-light hypothesis is not generally
satisfied \citep{tk04,hum10}. \citet{cap13} revisited the previous
analysis on a new set of galaxies with an expanded dataset of
spatially resolved kinematics, introducing different models with
luminous and dark components. They claim:
\begin{equation}
M(R_{\rm e})\approx {2.5R_{\rm e}\over G}\sigma^{2}_{\rm ap}(R_{\rm e})\ ,
\end{equation}
which would be essentially the same result as before if light traced
mass.

Analogous formulae have been derived for DM-dominated systems --
though the focus has been on dwarf spheroidal galaxies (dSphs), rather
than ellipticals. For a dSph with a Plummer luminosity profile and a
flat line of sight velocity dispersion $\sigma_{\rm p}$,
\citet{Wa09} showed that the mass within the effective radius is
\begin{equation}
M(R_{\rm e}) \approx {2.5R_{\rm e}\over G} \sigma^{2}_{\rm p} .
\label{eq:walk}
\end{equation}
In particular, \citet{Wa09} argued from Jeans solutions that the mass
within the half-light radius is robust against changes in the velocity
anisotropy and halo profiles. \citet{wol10} discovered a different,
but related, formula in which $\RM$ is the radius of the sphere
enclosing half of the total light $r_{1/2}$, whilst the velocity
dispersion is averaged over large radii
\begin{equation}
M(r_{1/2})\approx \frac{3r_{1/2}\langle\sigma^{2}_{p}\rangle_{\infty}}{G}\ .
\end{equation}
They provided a theoretical justification, based on the Jeans equations
under the hypothesis that the velocity dispersion profile is
approximately flat. \citet{amo11} extended this idea by looking for
masses robust against variation in the concentration and form of the
DM halo profile, using a particular class of distribution functions.
They advocated the formula
\begin{equation}
M(1.7R_{\rm e}) \approx {5.8R_{\rm e}\over G} \sigma^{2}_{\rm p}(R_{\rm e}) ,
\label{eq:amo}
\end{equation}
and so found that the mass enclosed within $r=1.7 R_{\rm e}$ was best
constrained.  A similar approach was pursued by \citet{chu10}; there,
the $\sigma_{\rm p}$ profiles of S{\'e}rsic tracers with a flat
rotation curve ($\gamma=2$) were studied, with particular emphasis on
isotropic, completely radial or completely tangential stellar orbits,
to identify the location where any dependence on anisotropy is
minimised. Using the assumption that the total density profile is
$\rho \sim r^{-2}$ enabled them to find fully analytical results.

All these formulae share a common ancestry, though they apply to
different luminosity profiles and dark halo laws. They all relate the
mass enclosed at a specific radius $\RM$ with the velocity
dispersion either at, or averaged within, a particular radius
$\RS$ based on different choices for the distribution function
of the stellar populations.  Here, we will show how the results of
Section 2 can be used systematically to construct mass estimators
tailored for elliptical galaxies with S{\'e}rsic profiles.

\begin{figure}
	\centering
        \includegraphics[width=0.45\textwidth]{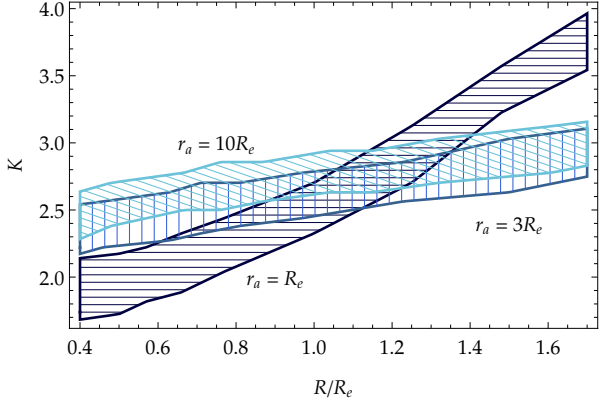}
\caption[Mass from the kinematic profile]{ \small{The coefficient
    $\Kv= v_{\rm c}^{2}(\RM)/ \sigma^{2}_{\rm p}(R)$ versus radius $R$
    for a de Vaucouleurs luminous profile in scale-free total
    densities, with Osipkov-Merritt anisotropy and radii $r_{\rm
      a}/R_{\rm e}=1,3,10$ (from the most to the least steep
    sequences). At each value of $R/R_{\rm e},$ a range is allowed for
    $\Kv$ corresponding to the freedom $1<\gamma<3.$ $\RS$ is the
    radius where the hatched zones intersect and so the dependence on
    anisotropy is minimised. The curves are computed using
    eq.~(\ref{eq:projsigma}).}  }
\label{fig:Kv}
\end{figure}

\subsection{Masses from the Kinematic Profiles}\label{sect:makinpr}

Without much loss of generality, we can operate within the framework
of scale-free total densities. In fact, the results of \citet{tk04},
\citet{mam05} and \citet{hum10}, which stem from analyses of different
tracers in different samples of early-type galaxies, suggest that a
realistic total density profile is scale-free to a first
approximation.  Then, each panel of Fig.~\ref{fig:vdispR} shows a
noteworthy property of the profiles $\sigma_{\rm p}(R),$ namely the
existence of a particular location $\RS$, where the dependence on the
exponent $\gamma$ is minimal.  Its value depends on the anisotropy
profile $\beta$ and on the circular velocity $v_{\rm c}$ at $R_{\rm
  e}.$ Also, the proportionality coefficient between $v_{\rm c}(R)$
and $\sigma_{\rm p}(R)$ varies between two extremes in the range
$1<\gamma<3$. We can synthesize this as:
\begin{equation}
v_{\rm c}^{2}(R_{\rm e})= K\sigma_{\rm
  p}^{2}(R_{\sigma}(\beta))\ ,
\end{equation}
where $K$ is a dimensionless constant, which may itself depend on the
anisotropy, as well as other dimensionless parameters.
\begin{table}
\begin{center}
\renewcommand{\tabcolsep}{0.2cm}
\renewcommand{\arraystretch}{0.5}
\begin{tabular}{| c | c | c | c |}
\multicolumn{4}{c}{} \\
	\hline 
	$n$ & $\RS/R_{\rm e}$ & $\RM/R_{\rm e}$ & $K$ \\  
	 &  &  & $\equiv v_{c}^{2}(\RM)/\sigma^{2}_{p}(R_{\sigma})$ \\  \hline
	1 & $0.81\pm0.07$ & $1.78\pm0.05$ & $3.03\pm0.37$ \\   
	2 & $0.97\pm0.10$ & $2.2\pm0.4$ & $2.95\pm0.35$ \\   
	3 & $1.12\pm0.12$ & $3.1\pm0.7$ & $2.86\pm0.25$ \\   
	4 & $1.15\pm0.15$ & $3.4\pm0.9$ & $2.78\pm0.15$ \\ 	  
	5 & $1.20\pm0.18$ & $3.9\pm1.1$ & $2.70\pm0.07$ \\  
	6 & $1.23\pm0.21$ & $4.33\pm1.33$ & $2.70\pm0.07$ \\ \hline
    \hline
\end{tabular}
\end{center}
\caption[Local Pinch radii]{The radii $\RS$ and $\RM$ in units
    of the effective radius and coefficient $K$ in eq.~(\ref{eq:Kdef})
  for different S{\'e}rsic indices $n$. The uncertainties are
  estimated by the excursion around the mid-value in plots
  analogous to Fig.~\ref{fig:Kv}}.
\label{tab:kpmest}
\end{table}

If a different radius $\RM$ is chosen as the one where $v_{\rm c}$ is
measured, the dependence $\RS$ on $\beta$ changes.  Then, we can seek
the radius $\RM$ such that the variation of $R_{\sigma}$ with $\beta$
is as small as possible. In this case, we obtain a relation of the
kind (\ref{eq:GMR}), where the radii $\RS$ and $\RM$ are the
ones where the measurements of velocity dispersion and enclosed mass
give the tightest excursion in the proportionality coefficient. In
other words, we are interested in finding a triplet $(\RS,\RM,K)$ such
that the relation
\begin{equation}
v_{\rm c}^{2}(\RM) = {GM(\RM)\over \RM} = K \sigma_{\rm p}^2 (\RS)
\label{eq:Kdef}
\end{equation}
holds with the smallest possible scatter over $\beta$ and $\gamma.$
 
Fig.~\ref{fig:Kv} shows the result of this strategy when $\Sigma(R)$
is a de Vaucouleurs profile with Osipkov-Merritt anisotropy laws. The
hatched zones intersect, and dependence on anisotropy minimised,
provided $\RS \approx1.2R_{\rm e}$ and $K \approx 2.8,$ which happens
when $\RM\approx 3R_{\rm e}$.  All these values are subject to mild
systematic uncertainty, estimated to be typically $\approx10\%$ from
Fig.~\ref{fig:Kv}.  Taking just tbe most probable values, we obtain
\begin{equation}
M(3.4R_{\rm e}) \approx {9.4 R_{\rm e}\over G} \sigma^{2}_{\rm
    p}\left(1.2R_{\rm e}\right).
    \label{eq:deVpinch}
\end{equation}
In other words, if the velocity dispersion of a de Vaucouleurs tracer
is measured at $\sim 1.2 R_{\rm e},$ then the mass just beyond
$3R_{\rm e}$ is well-constrained against variations in power-law index
$\gamma$ and anisotropy $\beta$. Note that if we further require that
light traces mass, then $M(3.4 R_{\rm e})$ is practically the total
mass and our result is equivalent to eq.~(\ref{eq:ill}) derived by
\citet{Ill76}. The roughly $10\%$ difference in the coefficients can
be ascribed to the choice of one particular mass model and variation
of $\sigma_{\rm p}$ with radius.

Our result can also be usefully compared with the work of
\citet[Section 5.2]{cou13}, who used the aperture average velocity
dispersion within $3R_{\rm e}$ and concluded that this was not
sufficient to constrain the enclosed mass at large radii.  Here, we
have shown that the line of sight velocity dispersion and shown that
it is surprisingly discriminating and provides a powerful way to study
the mass budget at large radii.

The same procedure can be repeated for other S{\'e}rsic-like profiles
of the surface brightness $\Sigma(R),$ {as summarised in Table
  \ref{tab:kpmest}.}  For example, in the case of an exponential law
$\Sigma(R)\propto \exp ({-1.67R/R_{\rm e}}),$ it yields
\begin{equation}
M(1.78 R_{\rm e}) \approx {4.8 R_{\rm e}\over G} \sigma^{2}_{\rm
    p}(0.81R_{\rm e}),
\end{equation}
\citep[cf][]{amo11}.  Though the coefficient $K$ and radius $\RS$ vary
weakly with the S{\'e}rsic index, the greatest variation is found in
the radius $\RM$, where the enclosed mass is estimated.

If we restrict to profiles with a nearly flat rotation curve, as
suggested from the weak homology arguments, the velocity dispersion
changes slowly with radius (cf Fig.~\ref{fig:vdispR}). Then, all the
radial dependence of enclosed mass is in $M(\RM)\propto \RM,$ as the
velocity dispersion $\sigma_{\rm p}$ is constant to good accuracy and
provides an overall mass normalisation. This is consistent with the
linear scaling $M\propto R$ from the density $\rho\propto r^{-2}$,
whilst the radius $\RM$ is simply a special point at which
uncertainties from anisotropy are minimised. However, the hypothesis
of weak homology comes with a significant \textit{caveat} that forbids
the restriction to $\gamma=2$ when examining single galaxies,
especially when $\RM$ is appreciably larger than the effective radius.

\label{sect:apmest}
\begin{figure}
	\centering
        \includegraphics[width=0.45\textwidth]{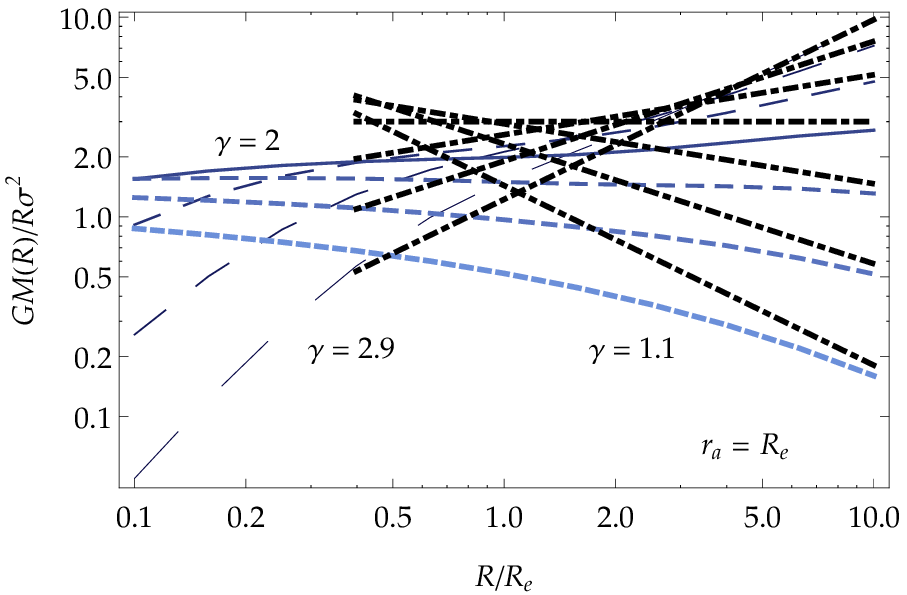}
        \includegraphics[width=0.45\textwidth]{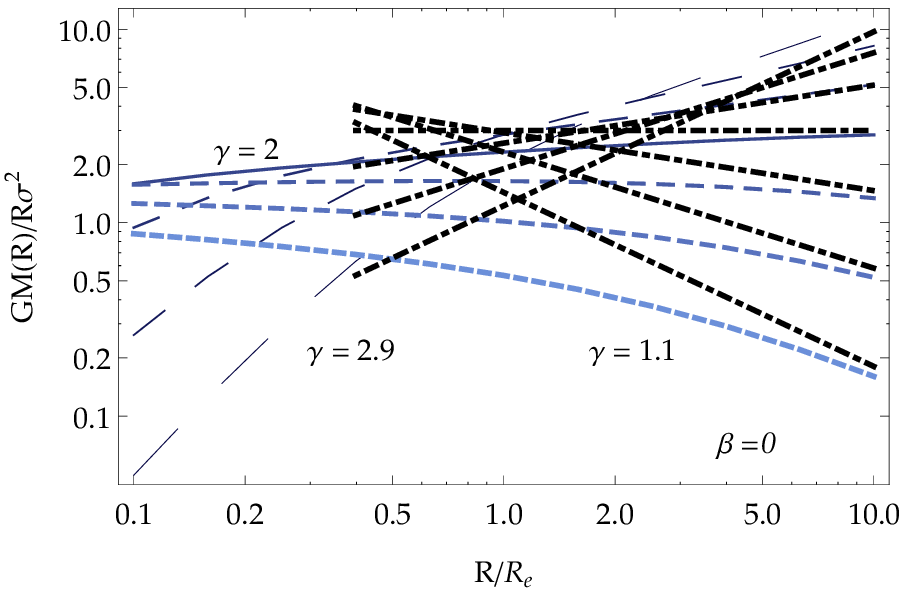}
\caption[Aperture masses]{ \small{Aperture mass estimators
    where $\sigma^{2}=\sigma^{2}_{\rm ap}(R)$ (dashed lines)
      or $\sigma^{2}_{\rm ap}(\infty)$ (dot-dashed curves). Again, the
      tracer has a de Vaucouleurs luminous profile with
      Osipkov-Merritt anisotropy profile $\beta(r)=r^{2}/(r^{2}+r_{\rm
        a}^{2})$ and power-law total density $\rho_{\rm tot}\propto
      r^{-\gamma}$. Top panel: $r_{\rm a}=R_{\rm e};$ bottom: $r_{\rm
        a}\gg R_{\rm e}$. The curves are computed via
      eq.~(\ref{eq:apsigma}).  The case $\gamma=2$ is marked with solid
      lines.  The dot-dashed curves corresponding to infinite aperture
      sizes are the same in both panels, but are plotted twice in
      order to ease the comparison with the cases with finite
      aperture.}}
\label{fig:apmass}
\end{figure}

\subsection{Aperture Masses and the Virial Limit}

Measuring the velocity dispersion at the exact location $R_{\sigma}$
is not possible in practice: the {observed} velocity dispersion is
always an average over some aperture, even when long-slit or
integral-field spectroscopy is performed. On the other hand, it often
happens that the radial average $\sigma^{2}_{\rm ap}(\RM)$ is
available. For example, fibre-averaged kinematics are usually measured
over typical lengths that are comparable to the effective radius, as
for example in the SLACS sample~\citep{aug10}.

This suggests another class of estimators, in which $\sigma^{2}_{\rm
  ap}(\RM)$ is used to measure the mass. As the dashed lines in
Fig.~\ref{fig:apmass} show, the sequences for $\sigma^{2}(R)/v_{\rm
  c}^{2}(\RM)$ still have an appreciable `pinch' at a special location
($R_{\sigma}\approx 0.5R_{\rm e}$) for a given anisotropy
radius. However, there is no analogue of the intersecting regions in
Fig.~\ref{fig:Kv} as the anisotropy varies.  The only exception is in
the virial limit, which is obtained by considering the average value
$\sigma^{2}_{\rm ap}(\infty)$ over the whole system.  It is well-known
that the virial theorem for spherical systems is independent of
anisotropy.  However, the aperture average over large radii is not
always available with acceptable accuracy, even for nearby galaxies. A
remarkable exception is given by the kinematics of resolved, extended
tracers like globular clusters and planetary nebulae orbiting around
the outer parts of nearby early-type galaxies (as discussed in Paper
II of this series).

\begin{figure}
	\centering
        \includegraphics[width=0.45\textwidth]{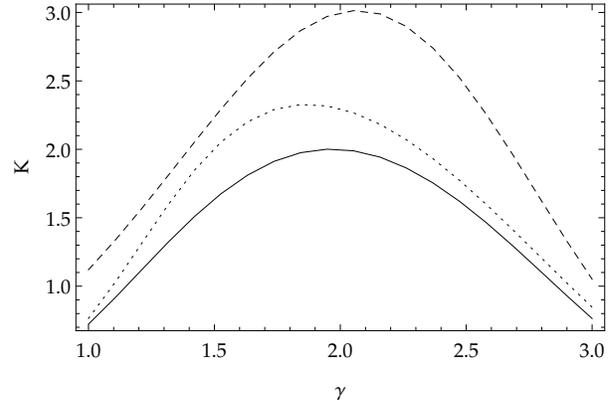}
\caption[Mass-model dependence of the mass estimator]{\small{ The
    aperture mass estimator at the radius $R_{\sigma}$ (cf
    Fig~\ref{fig:apmass}) for a de Vaucouleurs surface
      brightness with anisotropy profile from eq.~(\ref{eq:beta}),
      embedded in power-law total densities $\rho_{\rm tot}\propto
      r^{-\gamma}$.  Solid curve: $\beta_{\infty}=1$ and $r_{\rm
        a}=R_{\rm e};$ dotted line: $\beta=0;$ dashed line:
      large-aperture estimator (eq.~\ref{eq:mupl}).}  The curves are
    computed using the formulae in Section \ref{sect:apavkin} and
    Appendix B.}
\label{fig:do}
\end{figure}

The dot-dashed lines in Fig.~\ref{fig:apmass} show the ratio
$GM(R)/[R\sigma^{2}_{\rm ap}]$ in the virial limit, which of course
remains unchanged for different anisotropy profiles.  Again, the
luminous profile has a de Vaucouleurs form and resides in a power-law
total density.  For such systems, \citet{aae13} have already shown for
$\gamma$ within the physical interval $1<\gamma<3$
\begin{equation}
\mu(R) = \frac{G M(R)}{R\sigma^{2}_{\rm ap}(\infty)}=
\frac{3\sqrt{\pi}\Gamma(2-\gamma/2)}{2\Gamma\left((5-\gamma)/2\right)}\times\frac{R^{2-\gamma}}{\langle
  R^{2-\gamma}\rangle}\ ,
\label{eq:mupl}
\end{equation}
where angled brackets represent luminosity averages.  By studying the
dependence of $\mu(R)$ on $\gamma$, we find that
\begin{equation}
\mu(\RM)_{(\gamma=1)}=\mu(\RM)_{(\gamma\rightarrow3)}\ .
\end{equation}
This location $\RM$ can also be found analytically. In particular, if
the surface brightness is of the S{\'e}rsic form given in
eq~({\ref{eq:sers}), then
\begin{equation}
\RM=R_{\rm e}b_n^{-n}\sqrt{2\Gamma(3n)/\Gamma(n)}\ .
\label{eq:sweet}
\end{equation}
This implies that
 \begin{equation}
 \RM/R_{\rm e}\approx1.05\mathrm{dex}[-0.019(n-4)]
 \label{eq:mamon}
 \end{equation}
to $0.4\%$ relative accuracy when $1<n<10$, whence $\RM\approx R_{\rm
  e}$, as already suggested by Fig.~\ref{fig:apmass}.}

Having determined the radius that minimizes model dependence, we must
now assess the problem of systematics.  The coefficient for the virial
or large aperture limit takes the value $\Kv=3$ in the flat
rotation curve case, as shown in Fig.~\ref{fig:do}. It is somewhat
smaller for a finite radius aperture.  If we have no prior knowledge
on the density exponent, the coefficient $\Kv$ will be typically
distributed uniformly in $1\lesssim \Kv\lesssim 2.5$ and as
$(3-\Kv)^{-1/2}$ when $2.5\lesssim \Kv<3$. This follows from
approximating the dashed curve in Fig.~\ref{fig:do} by a parabola for
$\Kv \ge 2$ and straight line otherwise. The value $\Kv= 3$ is the
most likely, because $\mu(\RM)$ is approximately quadratic in $\gamma$
and always peaks near $\gamma=2$ (see Fig.~\ref{fig:do} and
eq.~\ref{eq:mupl}). However, the mean value of $\Kv$ for a uniform
prior on $\gamma$ is systematically lower than $3$. Its precise value
depends on the photometric profile through eq.~(\ref{eq:mupl}). For a de
Vaucouleurs profile, it is straightforward to establish from Monte
Carlo simulations that $\Kv \approx 2.3$.

By solving the Jeans equations and fitting kinematic profiles,
\citet{wol10} argued that $\Kv\approx3$ and $\RM=r_{1/2}\approx 1.3
R_{e}$ for a diverse set of systems. However, explicit
counter-examples are known for which the value $\Kv =3$ is never even
reached \citep[for instance, the models of][]{wil02}.  If the results
of \citet{gav07} and \citet{hum10} are valid in general for elliptical
galaxies, then the finding that $\Kv\approx3$ means that the total
density profile has $\gamma\approx 2$ in those systems near the
effective radius. The same remark holds here as in the case of
Cappellari et al. (2006): even if the mean behaviour is well fit by
$\gamma\approx 2$ near $R_{\rm e},$ individual variations from this
simple case are substantial.

If we are interested in learning about the density profile and mass
content in a particular galaxy, we cannot simply rely upon $\Kv\approx
3,$ as this would automatically bias our estimates towards a perfectly
flat rotation curve.  As a general rule, we advocate taking
$\Kv\approx 2.3,$ which follows from a uniform prior on $\gamma,$ and
 thus using the approximation
\begin{equation}
M(\RM)\approx {2.3 \RM\over G} \sigma^{2}_{\rm ap}(\infty)
\label{eq:globest}
\end{equation}
as a first estimate of the mass enclosed at the pinch radius in a
model-independent manner.  The radius $\RM$ for S{\'e}rsic profiles
does not vary substantially from $R_{\rm e}$.  This
formula~(\ref{eq:globest}) is valid provided $\sigma^{2}_{\rm
  ap}(\infty)$ is known, as this case for early-type galaxies with
extended populations of globular clusters and planetary nebulae.

\begin{figure}
	\centering
        \includegraphics[width=0.45\textwidth]{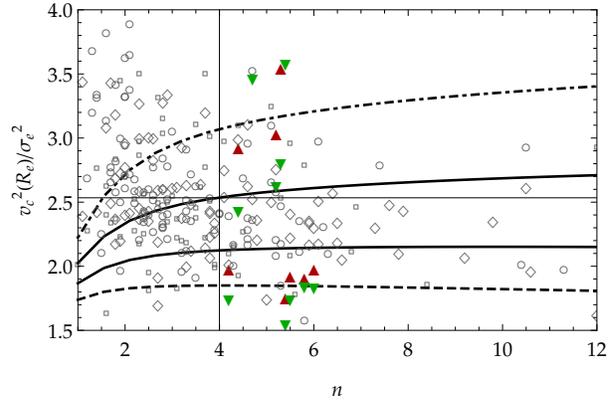}
\caption[Dependence with S{\'e}rsic index and anisotropy]{
  \small{Ratio of squared circular velocity at the effective radius,
    $v_{\rm c}^2 (R_{\rm e}),$ to the aperture-averaged velocity
    dispersion $\sigma^{2}_{\rm ap}(R_{\rm e})$ as a function of the
    S{\'e}rsic index of tracer. The solid curve corresponds to
    isotropy and $\gamma=2,$ the dashed line to
    $\gamma=2.1,\beta(R_{\rm e})=1/2$ and the dotted line to
    $\gamma=2.05,\beta(R_{\rm e})=1/4;$ the dot-dashed line shows the
    case $\gamma=2,\beta=-1/2.$ For an isotropic de Vaucouleurs tracer
    ($n=4$) and flat rotation curve, the ratio is approximately
    5/2. Values from \citet{per13} are shown as triangles, including
    (green, pointing downwards) or neglecting (red, upwards) aperture
    corrections.  Open symbols are values from \citet{cap13}, for
    galaxies where a S{\'e}rsic profile gives a good (circles),
    medium-quality (diamonds) or bad (small squares) photometric
    fit.}}
\label{fig:mestReff}
\end{figure}

\subsection{Finite Apertures}

A first general feature, already noticeable from Figs~\ref{fig:apmass}
and \ref{fig:do}, is that the model dependence is slightly smaller for
the finite radius estimator ($\sigma^{2}=\sigma^{2}_{\rm ap}(R)$) than
for the one with infinite radius.  This is because in the virial limit
the global average $\sigma^{2}_{\rm ap}(\infty)$ must be the same for
all possible anisotropies that correspond to acceptable solutions,
whence the larger variability.  Second, the mass estimator at fixed
$R$ and $\gamma$ generally has a lower value for the finite radius
choice
\footnote{The only exception is $\gamma<1.5$ and $R>1.5R_{e}$ (bottom
  panel of Fig.~\ref{fig:apmass}), i.e. shallow total density profiles
  and large apertures.}. This means that, if we assumed that
\textbf{the velocity dispersion profile is flat}, we would slightly
over-estimate the enclosed mass with respect to another equally
plausible choice, namely isotropy ($\beta=0$) at all radii
~\citep{aae13}.

Obtaining pinch radii and masses from finite apertures is harder, as
it is not possible to give general results unless additional
conditions are imposed. A simple mass estimator can be obtained by
invoking the weak homology hypothesis. For example, if we assume that
$\gamma = 2$ and $\beta=0,$ we readily obtain from eqs~(\ref{eq:iso})
and (\ref{eq:28})
\begin{equation}
\sigma^{2}_{\rm ap}(R_{\rm a})=\frac{GM(R_{\rm e})}{3R_{\rm e}}\times
\left(1+\frac{R_{\rm a}^{3}\int_{R_{\rm a}}^{\infty}\Sigma(s)s^{-2}\mathrm{d}s}
{\int_{0}^{R_{\rm a}}\Sigma(s)s\mathrm{d}s}\right),
\label{eq:finap}
\end{equation}
within the aperture radius $R_{\rm a}$. This formula is given by
\citet{chu10}, who also found complementary results for completely
radial ($\beta\rightarrow1$) or tangential ($\beta\rightarrow-\infty$)
orbits, still adopting $\gamma=2.$

When $\gamma\approx2$ and $\beta$ is small, a de Vaucouleurs surface
brightness leads to $\sigma^{2}_{\rm ap}(R_{\rm e})
\approx1.2\sigma^{2}_{\rm ap}(\infty).$ In this case, the enclosed
mass at radii $\RM \approx R_{\rm e}$ can be estimated by replacing
$\sigma^{2}_{\rm ap}(\infty)$ with $\sigma^{2}_{\rm ap}(\approx R_{\rm
  e})$ and $\RM$ with $R_{\rm e}$ in equation (\ref{eq:globest}),
provided the proportionality coefficient is adjusted to $\approx3/1.2=
5/2$. The mass from the finite-aperture sweetspot
(Fig.~\ref{fig:apmass}), linearly extrapolated to the effective
radius, would have a coefficient $K\approx2.4,$ which is halfway
between the large-aperture blind average and the weak homology
case. The ratio $v_{c}(R_{\rm a})/\sigma_{\rm ap}(R_{\rm a})$ between
circular velocity and average second moment within an aperture-radius
$R_{\rm a}$ depends weakly on $R_{\rm a}/R_{\rm e},$ as long as this
is around unity.

Then, a formula with $\RM\approx R_{\rm a}\approx R_{\rm e}$ and
$K\approx2.4$ is the simplest to use for early-type galaxies with
stellar velocity dispersion data largely confined to within one or two
effective radii, when the S{\'e}rsic index is close to $n=4.$

\subsection{Insights into Weak Homology}

Weak homology arguments are probably appropriate for nearby early-type
galaxies.  Fig.~\ref{fig:mestReff} shows the ratio $GM(R_{\rm
  e})/[R_{\rm e}\sigma^{2}_{\rm ap}(R_{\rm e})]$ for S{\'e}rsic
luminous components, as a function of the S{\'e}rsic index $n$ using
eq.~(\ref{eq:finap}) and $R_{\rm a}=R_{\rm e}.$ The dynamical analysis
of early-type galaxies by \citet{cap13} is summarised here by the open
symbols.  Regardless of the adequacy of the single S{\'e}rsic fit to
the photometric profile, which is indicated by different symbols, a
trend of $v_{\rm c}(R_{\rm e})/\sigma_{\rm ap}(R_{\rm e})$ with the
best-fitting S{\'e}rsic index $n$ is apparent. If the mass inference
is robust around $R_{\rm e},$ we can interpret this behaviour via
models with different anisotropy or power-law index.  In particular,
galaxies with lower (higher) $n$ have stars on slightly tangential
(radial) orbits on average. As shown in \citet{kra13}, nearby
early-type galaxies typically consist of bulge and disk components
with variable size and luminosity-ratios. If the bulge (or the disk)
dominates the photometric profile, that will drive the best-fitting
S{\'e}rsic index towards higher (or lower) values. Then, at least part
of the trend illustrated in Fig.~\ref{fig:mestReff} can be simply
understood as a variation of bulge-to-disk ratio, with disks (bulges)
having more stars on circular (radial) orbits.
 
Recently, \citet{per13} have cautioned against the approximation of
weak homology when compact massive galaxies, especially at higher
redshift, are examined. In their analysis, they find that dynamical
masses estimated as in Cappellari et al. (2006, 2013) imply negative
DM fractions. Equivalently, their inferred stellar masses can exceed
the dynamical estimates by almost an order of magnitude.

Since the mass within $R_{\rm e}$ is given by at least the luminous
component, we can consider $GM_{\star}/[2R_{\rm e}\sigma^{2}_{\rm
    ap}(R_{\rm e})]$ as a lower bound on $v_{\rm c}^{2}(R_{\rm
  e})/\sigma^{2}_{\rm ap}(R_{\rm e})$ and check how that compares with
the behaviour of nearby ellipticals.  The analysis in \citet{per13}
relies on stacked spectra to obtain velocity dispersions and stellar
masses, assuming a Salpeter IMF, in different redshift bins. At first
sight, their results seem hard to reconcile with diverse homology
arguments \citep{ber02,cap06,tay10,cap13}, or lensing results
\citep{nip08}. However, the velocity dispersion should be averaged
within the effective radius, in order to operate a fair
comparison. When the simple correction $\sigma_{\rm ap}(R)\propto
R^{-0.066}$ is made (c.f. Section 2.3), most of the objects fall back
into the range spanned by weak homology. This is merely a consistency
check, since applying the same kind of aperture correction to each
galaxy tacitly assumes some kind of homology across the sample. The
discrepancy is still present for the most compact ones, which may then
be interpreted as a set of fast rotators. Spatially resolved kinematic
information will tell if this is the case.  Also, the choice of IMF
may play a role. When dynamical masses are inferred via gravitational
lensing, then a (universal) Salpeter IMF implies negative DM fractions
for some of the SLACS galaxies~\citep{aug10}. Interestingly, there is
evidence to suggest a dichotomy in early-type galaxies. Slow rotators
show a tendency towards a Salpeter IMF, and fast rotators towards a
Chabrier IMF \citep{gri09,aug10,ems11,suy12}. Moreover, the IMF is
known to vary with velocity dispersion \citep{cap12,spi13}. The
resolution of the problem indicated by \citet{per13} may be that both
a non-universal IMF and more detailed kinematic information are
required when dealing with compact massive galaxies at higher
redshift, although part of the tension is already alleviated when
aperture corrections are included.

\section{Discussion and Conclusions}

We have shown how, under the approximation of spherical symmetry, the
line-of-sight velocity dispersion can be computed by means of
quadratures involving the surface brightness profile $\Sigma(R)$ and a
kernel that depends on the mass model and on the anisotropy. This
avoids the need for explicit de-projection of the surface brightness
to give the luminosity density, subsequent solution of the Jeans
equations and final re-projection to give the line of sight
dispersion.  We have provided simple approximations for the kinematics
at large distances or mild anisotropy.

The results on kinematic profiles can be adapted to include the
process of averaging through circular apertures of varying
size. Results for other cases (long-slit measurements, averages
through an annulus, point-spread-function blurring) can be obtained by
simple combinations of the ones for a circular aperture.  The
aperture-averaged velocity dispersion can be computed by means of
single integral over the stellar density profile modulated by a kernel
encoding the dependence on mass and anisotropy. If the surface
brightness $\Sigma(R)$ is used, the quadratures are (at worst) double
integrals and the kernels can be re-written as combinations of special
functions. For some special cases (including constant anisotropy with
$\beta_{\infty}=1,1/2$ and scale-free total densities), the kernel can
be written explicitly in terms of elementary functions.

The aperture-averaged kinematic profiles for a de Vaucouleurs luminous
component in scale-free total densities ($\rho_{\rm tot} \propto
r^{-\gamma}$) reproduce the empirical behaviour observed in over 25
early-types in the SAURON survey~\citep{cap06}, provided the density
exponent is $\gamma=2.05\pm0.05$ and anisotropy at the effective
radius is mild ($0\leq\beta(R_{\rm e})\lesssim0.5$). This result
agrees with the findings of \citet{koo09}, which are based on the
analysis of 58 lensing galaxies in the SLACS sample \citep{bol06}. At
least as regards bulk properties, elliptical galaxies are seemingly
well-represented by the simple isotropic models with a flat rotation
curve.

Mass estimators can be derived by examining the kinematic profiles or
aperture-averaged velocity dispersions.  When the surface brightness
$\Sigma(R)$ is measured with sufficient accuracy, one strategy is to
determine the location $\RM$ within which the enclosed
mass is best constrained and the radius $R_{\sigma}$ at which
kinematics should be measured in order to produce the tightest mass
estimate.  In the more common case of aperture-averaged kinematics, we
have not found simple estimators for a de Vaucouleurs profile in
scale-free total density that are truly robust against changes in
anisotropy, except in the large aperture or virial limit.

For extended tracers in the outer parts of elliptical galaxies, such
as globular clusters or planetary nebulae, the velocity dispersion
averaged over a large aperture is in principle measurable.  So,
eq.~(\ref{eq:globest}) provides a simple estimate of the mass enclosed
at a radius $\RM$ that, for a de Vaucouleurs profile, is
near to the effective radius. More commonly, the kinematical
information is available only for populations within an effective
radius or so.  Then we advocate using
\begin{equation}
M(R_{\rm e})\approx {2.4 R_{\rm e}\over G} \sigma^{2}_{\rm ap}
\label{eq:final}
\end{equation}
as the simplest mass-estimator in the absence of more detailed
information, provided the photometric profile is bulge-dominated (that
is, has a Sersic index $n\gtrsim 3.5$).  This is broadly consistent
with the estimator of Cappellari et al. (2006, 2013), namely that the
mass enclosed near the half-light radius is $M_{1/2}\approx 2.5R_{\rm
  e}\sigma^{2}_{\rm ap}(R_{\rm e})/G,$ even if we have derived the
result under completely different and more general hypotheses.  The
total mass enclosed within the effective radius appears to be a robust
quantity for S{\'e}rsic-like luminous profiles, independently of the
underlying mass model.

Our conclusions here are primarily theoretical. In a companion paper,
we put the machinery to work in an analysis of the globular clusters
of M87, and its implications for the mass distribution and orbits.

\section*{Acknowledgments}
AA thanks the Science and Technology Facility Council and the Isaac
Newton Trust for financial support. Discussions with Luca Ciotti,
Vasily Belokurov and Michele Cappellari are gratefully
acknowledged. We thank the referee, Gary Mamon, for careful and
patient readings of the manuscript, which helped improve it
significantly, and for suggesting eq.~(\ref{eq:mamon}).

\onecolumn

\begin{appendix}

\section{Mathematical Details}

Here, we give some of the technical details of the proofs required to
derive the formula in the main body of the paper.

\subsection{Proof of Equations (9) and (10)}
For any function $f(x,R)$, integration by parts gives
\begin{equation}
\int_{R}^{y}\frac{x f(x,R)}{\sqrt{y^{2}-x^{2}}}\mathrm{d}x=f(R,R)+
\int_{R}^{y}\partial_{x}\left(f(x)\right)\sqrt{y^{2}-x^{2}}\mathrm{d}x\ .
\end{equation}
Assuming that $f(R,R)$ vanishes and the integrals are uniformly
convergent, then we can differentiate the above with respect to $y$ to
obtain eq. (9). For eq. (10), we note that:
\begin{equation}
\int_{R}^{\infty}\nu(r)u(r,R)\mathrm{d}r = -\frac{1}{\pi}\int_{R}^{\infty}u(r,R)\int_{r}^{\infty}\frac{\Sigma^\prime(y)}{\sqrt{y^{2}-r^{2}}} \mathrm{d}y\mathrm{d}r
 = -\frac{1}{\pi}\int_{R}^{\infty}\Sigma^\prime(y)
\int_{R}^{y}\frac{u(r,R)}{\sqrt{y^{2}-r^{2}}}\mathrm{d}r\mathrm{d}y\ ,
\end{equation}
where primes denote differentiation. Here, we again assume that
$u(r,R)$ vanishes at $r=R$ and all the integrals are well defined,
Integrating by parts in $y$ and using eq. (9) with $f(r)=u(r,R)/r$,
then eq. (10) follows if we set
$u(r,R)=M(r)\left[\sqrt{r^{2}-R^{2}}+k_{\beta}(R,r)\right]/r^{2}.$

\subsection{Proof of Equations (15) and (17)}

Let us define $F(r)=GM(r)/r^{2}$ for conciseness. We start directly
from eq.~(4), multiply by $2\pi R,$ integrate in $0<R<R_{\rm a}$ and
reverse orders of integration between $R$ and $r:$
 \begin{eqnarray}
 L(R_{\rm a})\sigma^{2}_{\rm ap}(R_{\rm a}) &=& 4\pi
 \int_{0}^{R_{\rm a}}R\int_{R}^{\infty}\left(1-\beta(r)\frac{R^{2}}{r^{2}}\right)
 \frac{r}{\sqrt{r^{2}-R^{2}}}\int_{r}^{\infty}F(s)J_{\beta}(r,s)\mathrm{d}s\mathrm{d}r\mathrm{d}R\\
 \nonumber &=& 4\pi \int_{0}^{R_{\rm a}}
 \left[\int_{0}^{r}R\left(1-\beta(r)\frac{R^{2}}{r^{2}}\right)
 \frac{r}{\sqrt{r^{2}-R^{2}}}\mathrm{d}R\right]
 \int_{r}^{\infty}F(s)J_{\beta}(r,s)\mathrm{d}s\mathrm{d}r\\
 \nonumber &+& 4\pi\int_{R_{\rm a}}^{\infty}
 \left[\int_{0}^{R_{\rm a}}R\left(1-\beta(r)\frac{R^{2}}{r^{2}}\right)
 \frac{r}{\sqrt{r^{2}-R^{2}}}\mathrm{d}R\right]
 \int_{r}^{\infty}F(s)J_{\beta}(r,s)\mathrm{d}s\mathrm{d}r\ .
 \end{eqnarray}
The integrals in $R$ are easily performed and lead to
\begin{eqnarray}
  L(R_{\rm a})\sigma^{2}_{\rm ap}(R_{\rm a}) &=& 4\pi
  \int_{0}^{\infty}r^{2}(1-\frac{2}{3}\beta(r))
  \int_{r}^{\infty}F(s)J_{\beta}(r,s)\mathrm{d}s\mathrm{d}r\\
  \nonumber &-& 4\pi\int_{R_{\rm a}}^{\infty}
  \left(r\sqrt{r^{2}-R_{\rm a}^{2}}-\frac{2}{3}\beta(r)\frac{(r^{2}-R_{\rm a}^{2})^{3/2}}{r}\right)
  \int_{r}^{\infty}F(s)J_{\beta}(r,s)\mathrm{d}s\mathrm{d}r\\
  \nonumber &+& 4\pi R_{\rm a}^{2}
  \int_{R_{\rm a}}^{\infty}\frac{\beta(r)}{r}\sqrt{r^{2}-R_{\rm a}^{2}}
  \int_{r}^{\infty}F(s)J_{\beta}(r,s)\mathrm{d}s\mathrm{d}r\ .
\end{eqnarray}
The last line gives the third term in eq.~(15), provided we
exchange orders of integration between $r$ and $s.$ For the other two
terms, we also observe that
$\partial_{r}J_{\beta}(r,s)=-2\beta(r)J_{\beta}(r,s)/r$ and
$J(s,s)=J(r,r)=1,$ so that
\begin{eqnarray}
\int_{0}^{s}r^{2}J_{\beta}(r,s)\mathrm{d}r &=& \frac{1}{3}s^{3}
+\frac{2}{3}\int_{0}^{s}\beta(r)r^{2}J_{\beta}(r,s)\mathrm{d}r\ ,\\
\int_{R_{\rm a}}^{s}r\sqrt{r^{2}-R_{\rm a}^2}J_{\beta}(r,s)\mathrm{d}r &=&
 \frac{1}{3}(s^{2}-R_{\rm a}^2)^{3/2}
 +\frac{2}{3}\int_{R_{\rm a}}^{s}\beta(r)J_{\beta}(r,s)\frac{(r^{2}-R^{2}_{\rm a})^{3/2}}{r}\mathrm{d}r\ ,
\end{eqnarray}
whence eq.~(15), whose first line is obtained via
$\partial_{r}M(r)=4\pi\rho_{\rm tot}(r)r^{2}.$ Eq.~(17) follows by
Abel deprojection of $\nu$ and the same line of reasoning that led to
eq.~(10).

\subsection{Proof of Equations (20) and (24)}

When $\beta$ or $(s-r)/r$ are small, we may Taylor expand eq.~(3) to
obtain
\begin{equation}
J_{\beta}(r,s)\ \sim\ 1+2\int_{r}^{s}\beta(u)\mathrm{d}u/u.
\end{equation}
Then, we can approximate $J_{\beta}\sim 1$ in the integrals
$k_{\beta}(R,x)$ and $Z_{\beta}(R,x),$ to obtain first order
approximations in $|\beta|$ and $x-R.$ For higher order terms, the
whole behaviour of $\beta$ is necessary.  Eq. (20) is valid in
general, whereas eq. (24) is obtained in the limit $R\gg r_{\rm
  a},$ i.e. $\beta\sim\beta_{\infty}.$ An expansion accounting for other
terms in $r_{\rm a}/R$ is
\begin{equation}
k_{\beta}(R,x)\ \sim\ -\beta(R)(x^{2}-R^{2})^{1/2}
+\beta_{\infty}\frac{(1-\tfrac{2}{3}(\beta_{\infty}-r_{\rm a}^{2}/R^{2}))}
{(1+r_{\rm a}^{2}/R^{2})^{2}}(x^{2}/R^{2}-1)^{3/2}R+\mathcal{O}(\beta_{\infty}(x^{2}/R^{2}-1)^{5/2})R\ .
\end{equation}  
When $\Sigma$ decays sufficiently fast, higher-order terms are
suppressed and we obtain the asymptotic expressions
 \begin{equation}
 k_{\beta}(R,x)\sim -\beta(R)(x^{2}-R^{2})^{1/2}\ ,
 \end{equation}
 \begin{equation}
 Z_{\beta}(R,y)\sim \frac{1}{3R^{2}}(y^{2}-R^{2})^{3/2}.
 \end{equation}
These are usually sufficient to approximate $\sigma_{p}$ and $\sigma_{\rm
  ap}.$ The main exception is the case $\beta\rightarrow 1,$ when the
first non-trivial term in $\sqrt{x^{2}-R^{2}}+k_{\beta}(R,x)$ is
proportional to $(x^{2}-R^{2})^{3/2}.$

\subsection{Proof of Equations (21), (25) and (28)}

We start by noting that
\begin{equation}
\partial_{r}\left(r^{-\gamma}(r^{2}-R^{2})^{j/2}\right)\ =\ r^{-\gamma-1}(r^{2}-R^{2})^{j/2-1}
\left[(j-\gamma)r^{2}+\gamma R^{2}\right]\ .
\label{eq:useful}
\end{equation}
If $\rho_{\rm tot}=\rho_{0}(r/r_{0})^{-\gamma},$ then from eq. (10):
\begin{equation}
\Sigma\sigma^{2}_{p}(R)=\frac{8 G\rho_{0}r_{0}^{\gamma}}{3-\gamma}
\int_{R}^{\infty}y\Sigma(y)\int_{R}^{y}
\frac{\partial_{r}\left(r^{-\gamma}(\sqrt{r^{2}-R^{2}}+k_{\beta}(R,r)\right)}
{\sqrt{y^{2}-r^{2}}}\mathrm{d}r\mathrm{d}y\ .
\end{equation}
Now, eq. (21) (resp. 25) follows by exploiting equation (\ref{eq:useful})
 and eq. (20) (resp. 24), via the replacements $R=xy$ and $r=\sqrt{t}y.$
 
An analogous argument can be followed to obtain the average velocity
dispersion within a circular aperture. However, an alternative
procedure leads to more convenient formulae such as eqs.~(28) and
(29). We start by recasting eq.~(17) as
\begin{equation}
\sigma^{2}_{\rm ap}(R)\ =\ \frac{4G}{3L(R)}\left[I(0)-I(R)\right]\ ,
\end{equation}
where
\begin{equation}
\nonumber I(R)\ \equiv\ \int_{R}^{\infty}\Sigma(s)s\int_{R}^{s}
 \frac{\partial_{r}\left(M(r)(r^{2}-R^{2})^{3/2}/r^{3}\right)}{\sqrt{s^{2}-r^{2}}}
  \mathrm{d}r\mathrm{d}s
	\ =\ \int_{R}^{\infty}\Sigma(s)\frac{\mathrm{d}}{\mathrm{d}s}\int_{R}^{s}
	\frac{M(r)(r^{2}-R^{2})^{3/2}/r^{2}}{\sqrt{s^{2}-r^{2}}}\mathrm{d}r\mathrm{d}s\ .
\end{equation}
For a power-law total density $\rho_{\rm
  tot}(r)=\rho_{0}(r/r_{0})^{-\gamma},$ we have
\begin{equation}
I(R)\ =\ \frac{4\pi\rho_{0}r_{0}^{\gamma}}{3-\gamma}\int_{R}^{\infty}\Sigma(R)
 \frac{\mathrm{d}}{\mathrm{d}s}\int_{R}^{s}\frac{r^{1-\gamma}(r^{2}-R^{2})^{3/2}}{\sqrt{s^{2}-r^{2}}}\mathrm{d}r\mathrm{d}s\ .
\end{equation}
The derivative with respect to $s$ is:
\begin{eqnarray}
\nonumber \frac{\mathrm{d}}{\mathrm{d}s}\int_{R}^{s}\frac{r^{1-\gamma}(r^{2}-R^{2})^{3/2}}{\sqrt{s^{2}-r^{2}}}\mathrm{d}r &=& \frac{\mathrm{d}}{\mathrm{d}s}\left(
  s^{4-\gamma}\int_{R/s}^{1}\frac{u^{1-\gamma}(u^{2}-(R/s)^{2})^{3/2}}{\sqrt{1-u^{2}}}\mathrm{d}u\right)\\
   &=& (4-\gamma)s^{3-\gamma}\int_{R/s}^{1}\frac{u^{1-\gamma}(u^{2}-(R/s)^{2})^{3/2}}{\sqrt{1-u^{2}}}
   \mathrm{d}u +3s^{3-\gamma}\left(\frac{R}{s}\right)^{2}\int_{R/s}^{1}
   \frac{u^{1-\gamma}\sqrt{u^{2}-(R/s)^{2}}}{\sqrt{1-u^{2}}}\mathrm{d}u\ .
   \label{eq:above}
\end{eqnarray}
Eq. (28) then follows by using
$(u^{2}-(R/s)^{2})^{3/2}=(u^{2}-(R/s)^{2})\sqrt{u^{2}-(R/s)^{2}},$
splitting the first integral in equation (\ref{eq:above}) and summing
the two terms proportional to $(R/s)^{2}.$ 

\section{Special Cases}

\subsection{Anisotropy Profiles with Analytic Kernels}

Here, we list some special cases of the kernels $k_\beta(R,x)$ defined
in eq.~(\ref{eq:kpdefn}) and $Z_\beta(R,y)$ defined in
eq.~(\ref{eq:apavps}) We recollect that these kernels are needed in
the quadratures for the line of sight and aperture-averaged velocity
dispersions respectively.

For the anisotropy profile (\ref{eq:beta}), the kernel $k_\beta$ can
be expressed in terms of hypergeometric functions, as indicated in
eq.~(\ref{eq:hyperg}). The corresponding result for $Z_\beta$ was not
given in the main text, and so we report it here
\begin{equation}
Z_\beta(R,y)
 =
\frac{\beta_{\infty}}{(4\beta_{\infty}^{2}-1)\sqrt{y^{2}-R^{2}}} 
\left[(r_{\rm a}^{2}+R^{2})_{2}F_{1}\left(1,-\beta_{\infty}\!-\!\frac{1}{2},\frac{1}{2},z\right)
\!-\!(y^{2}\!+\!r_{\rm a}^{2})\!-2\!\beta_{\infty}(y^{2}\!-\!R^{2})\right]\ ,
\end{equation}
where we have put $z=(R^{2}-y^{2})/(r_{\rm a}^{2}+R^{2})$. The kernel
is regular at $\beta_{\infty}=\tfrac{1}{2}$ and $y=R$, as may be
confirmed by careful Taylor expansion.

Some special cases reduce to elementary functions, and we briefly note
these results here.  In the Osipkov-Merritt case $\beta_{\infty}=1$,
we have
\begin{eqnarray}
k_{\beta}(R,x)&=&\frac{1}{2(r_{\rm a}^{2}+R^{2})^{3/2}}\left[(2r_{\rm a}^{2}+R^{2})(r_{\rm a}^{2}+x^{2})\mathrm{arctan}\sqrt{\frac{x^{2}-R^{2}}{r_{\rm
        a}^{2}+R^{2}}} -(2r_{\rm
    a}^{2}+3R^{2})\sqrt{(x^{2}-R^{2})(r_{\rm
      a}^{2}+R^{2})}\right],\\
Z_{\beta}(R,y) &=& \frac{1}{2\sqrt{r_{\rm a}^{2}+R^{2}}} 
 \left[(r_{\rm a}^{2}+y^{2})\mathrm{arcsin}\sqrt{\frac{y^{2}-R^{2}}{y^{2}+r_{\rm a}^{2}}}-\sqrt{(y^{2}-R^{2})(R^{2}+r_{\rm a}^{2})}\right]\ .
\end{eqnarray}
When $\beta_{\infty}=\tfrac{1}{2}$, we have
\begin{eqnarray}
k_{\beta}(R,x) &=&\frac{\sqrt{r_{\rm a}^{2}+x^{2}}}{2(r_{\rm a}^{2}+R^{2})}\left[2(r_{\rm a}^{2}+R^{2})\mathrm{arcsinh}\sqrt{\frac{x^{2}-R^{2}}{R^{2}+r_{\rm a}^{2}}}
-(2r_{\rm a}^{2}+ 3R^{2})\sqrt{\frac{x^{2}-R^{2}}{r_{\rm a}^{2}+x^{2}}}\right],\\
Z_{\beta}(R,y) &=& \frac{\sqrt{r_{\rm a}^{2}+y^{2}}}{2}
\left(\mathrm{arcsinh}\sqrt{\frac{y^{2}-R^{2}}{r_{\rm a}^{2}+R^{2}}}-\sqrt{\frac{y^{2}-R^{2}}{r_{\rm a}^{2}+y^{2}}}\right).
\end{eqnarray}
When $r_{\rm a}=0$, the models have constant anisotropy $\beta_\infty$ and
we obtain
\begin{eqnarray}
k_{\beta }(R,x) &=&\beta_{\infty}R(x/R)^{2\beta_{\infty}}\left[B\left(\beta_{\infty}-\frac{1}{2},\frac{1}{2}\right)-B\left(\frac{R^{2}}{x^{2}},\beta_{\infty}-\frac{1}{2},\frac{1}{2}\right)
+\frac{3}{2}B\left(\frac{R^{2}}{x^{2}},\beta_{\infty}+\frac{1}{2},\frac{1}{2}\right)-\frac{3}{2}B\left(\beta_{\infty}+\frac{1}{2},\frac{1}{2}\right)\right],\\
Z_{\beta}(R,y) &=&
\frac{\beta_{\infty}}{2}R(y/R)^{2\beta_{\infty}}\left[B\left(\frac{3}{2},\beta_{\infty}-\frac{1}{2}\right)-B\left(\frac{R^{2}}{y^{2}},\beta_{\infty}-\frac{1}{2},\frac{3}{2}\right)\right],
\end{eqnarray}
where $B(z,a,b)$ is the incomplete Beta function and
$B(a,b)=B(0,a,b)$. We note that equivalent formulae for the kernel
$k_\beta$ in the Osipkov-Merrit and constant anisotropy cases have
previously been given by \citet{mam05}.

\subsection{Large Radii and Small Anisotropies}
At large radii and/or small anisotropies, the line of sight velocity
dispersion can be written more conveniently:
\begin{eqnarray}
\nonumber \frac{\Sigma(R)\sigma^{2}_{\rm p}(R)}{1-\beta(R)} &\sim&
8G\int_{R}^{\infty}\Sigma(y)y\int_{R}^{y}\rho_{\rm tot}(x)
\frac{\sqrt{x^{2}-R^{2}}}{x^{3}\sqrt{y^{2}-x^{2}}}
\mathrm{d}x\mathrm{d}y + 8GR^{-3}\left(\int_{0}^{R}\rho_{\rm
  tot}(x)x^{2}\mathrm{d}x\right)\int_{R}^{\infty}\Sigma(y)yA(1,y/R)\mathrm{d}y\\ &+&
8GR^{-3}\int_{R}^{\infty}\Sigma(y)y\int_{R}^{y}\rho_{\rm
  tot}(x)x^{2}A(x/R,y/R)\mathrm{d}x\mathrm{d}y\ ,
\label{eq:A}
\end{eqnarray}
where the integral
\begin{equation}
A(\chi,\xi)=\int_{\chi}^{\xi}\frac{3-2r^{2}}{\sqrt{(\xi^{2}-r^{2})(r^{2}-\chi^{2})}}\frac{\mathrm{d}r}{r^{4}}
\end{equation}
does not depend on any mass model and can be tabulated separately.

Similarly, for aperture-averaged dispersions, when anisotropy is
sufficiently small, we have
\begin{eqnarray}
\nonumber \frac{3L(R)\sigma^{2}_{\rm ap}(R)}{16\pi G} &\sim& \int_{0}^{\infty}\Sigma(y)y\int_{0}^{y}\frac{\rho_{\rm tot}(x)x^{2}\mathrm{d}x}{\sqrt{y^{2}-x^{2}}}- 3(1-\beta(R))R^{-1}\left(\int_{0}^{R}\rho_{\rm tot}(x)x^{2}\mathrm{d}x\right)\int_{R}^{\infty}\Sigma(y)yB(1,y/R)\mathrm{d}y\\
 &-& 3(1-\beta(R))R^{-1}\int_{R}^{\infty}\Sigma(y)y\int_{R}^{y}\rho_{\rm tot}(x)x^{2}B(x/R,y/R)\mathrm{d}x\mathrm{d}y\ ,
\label{eq:B}
\end{eqnarray}
where again
\begin{equation}
B(\chi,\xi)=\int_{\chi}^{\xi}\sqrt{\frac{r^{2}-\chi^{2}}{\xi^{2}-r^{2}}}\frac{\mathrm{d}r}{r^{4}}\ .
\end{equation}
is independent of any model adopted and can be tabulated separately.

\end{appendix}

\label{lastpage}

\end{document}